# Towards Refactoring of GIPSY and DMARF Case Studies

## Project Milestone 4

Osama Yawar
Master of Applied Computer Science
Concordia University
Montreal, Canada
osamayawar@hotmail.com

Tahir Ayoub
Master of Software Engineering
Concordia University
Montreal, Canada
tahirayoub@live.com

Yashwanth Beeravelli
Master of Software Engineering
Concordia University
Montreal, Canada
yashwanthbeeravelli@gmail.com

Muhammad Nadir
Master of Software Engineering
Concordia University
Montreal, Canada
mu_nadir@encs.concordia.ca

Shahroze Jamil
Master of Software Engineering
Concordia University
Montreal, Canada
shahroze.srz@gmail.com

Parham Darbandi
Master of Software Engineering
Concordia University
Montreal, Canada
darbandi@gmail.com

## I. INTRODUCTION

The intent of this report is to do a background study of the two given OSS case study systems namely GIPSY and DMARF. It is a wide research area in which different studies are being carried out to get the most out of it. It begins with a formal introduction of the two systems and advance with the complex architecture of both.

GIPSY (General Intensional Programming System) is a multi-intensional programming system that delivers as a framework for compiling and executing programs written in Lucid Programming Languages. DMARF (Distributed Modular Audio Recognition Framework) is a Java based research platform that acts as a library in applications.

As these systems are in their evolving phase and a lot of research is being done upon these topics, it gives us motivation to be a part of this research carried out by Professor Serguei A. Mokhov and other known researchers to get a deeper look into the architectures of both the systems.

For the evaluation of quality of metrics of the two open source systems, we have used a tool namely Logiscope. It is a tool to automate the code reviews by providing information based on software metrics and graphs.

In this paper, Section II represents the OSS case studies of GIPSY and DMARF. All the summaries of team members were gathered together and defined a decent background for both GIPSY and DMARF separately and participates in presenting the metrics definition. The Metrics presented are of the DMARF and GIPSY measured using Logiscope. Table 3 in Section II.C outlines the individual selection of OSS case study papers. Section IV concludes the topic by identifying the benefits of both the tools. In Section V code smells are identified in source codes of DMARF and GIPSY and refactorings for those code smells are also suggested. Section VI refers to identification of design patterns in source code and Section VII discusses about how the refactoring is being implemented in projects. Finally, references and citations were mentioned at the end after a short glossary.

## II. BACKGROUND

### A. OSS Case Studies

#### 1) MARF

MARF (Modular Audio Recognition Framework) is an open-source project for the purpose of Audio recognition [1][18][2] also it has some algorithms which can be used for this sort of purpose. It has a recognition pipeline having a series of sequential steps -- sample loading, preprocessing, feature extraction, training/classification.

MARF is purely sequential with even little or no concurrency when processing a bulk of voice samples. Thus, the purpose of this work is to make the pipeline distributed and run on a cluster or a just a set of distinct computers to compare with the traditional version and do a thorough software engineering design for disaster recovery and service replication, communication technology independence, and the like. There are several distributed services, which some are more general, and some are more specific [19] [2].

*SYSTEM OVERVIEW*

Architectural Strategies
The main principles are [19]:

- Platform-Independence
- Database-Independent API
- Communication Technology Independence
- Reasonable Efficiency
- Simplicity and Maintainability
- Architectural Consistency
- Separation of Concern

MARF has different stages in it which are communicating each other with the help of pipelines, using these pipelines MARF stages communicated with each other in order to process data in a chained manner. These pipelines consist of four basic stages: Sample Loading, preprocessing, feature extraction and training/classification [2].

### 2) DMARF

The MARF was extended to DMARF as shown in Figure 1.

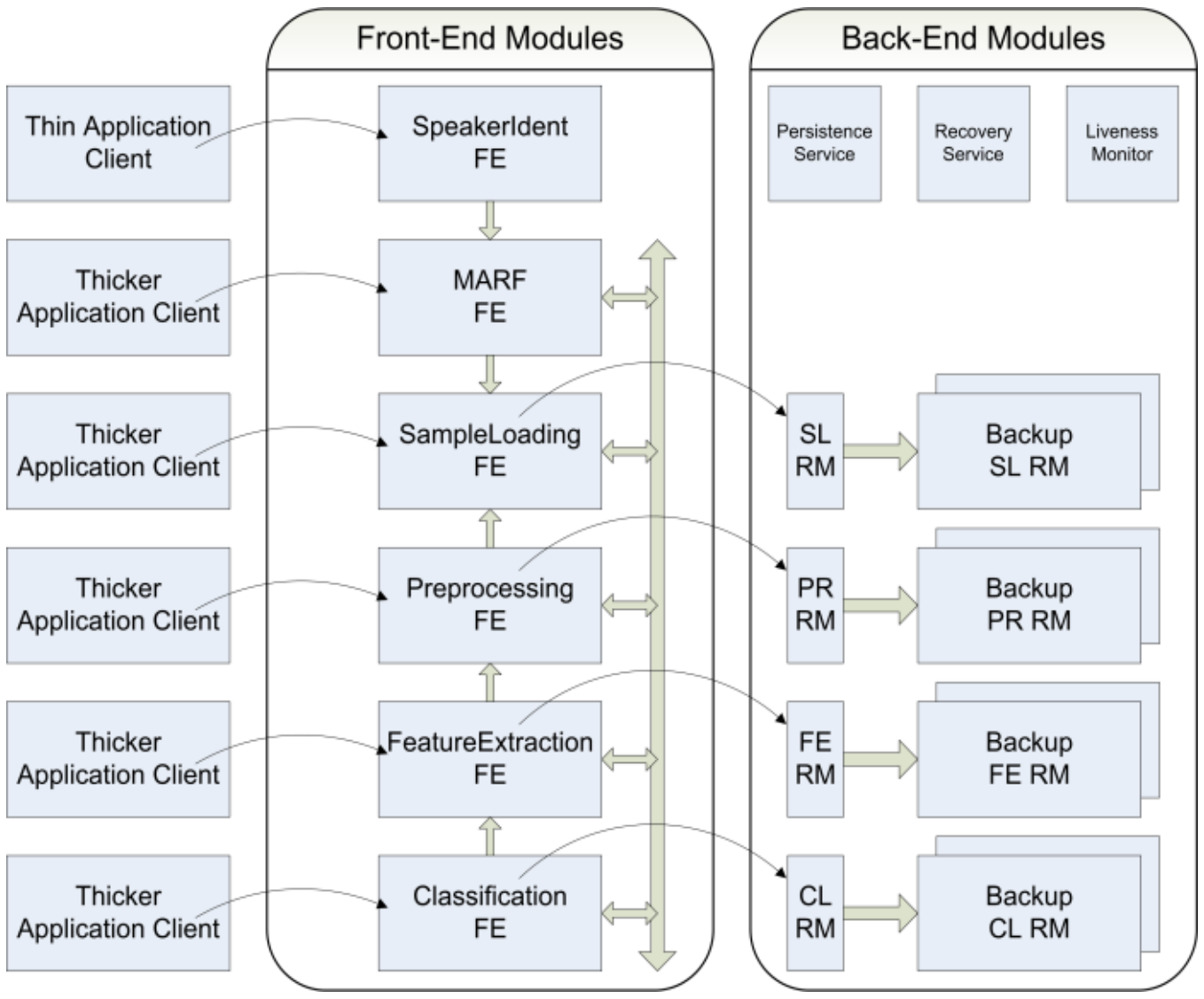

*Figure 1: Distributed MARF Pipeline*

The autonomic functioning of the distributed pattern-recognition pipeline plus its optimization is now primarily covered by DMARF as an autonomic system. The two major functionalities of DMARF as discussed in paper are: Training set classification data replication and Dynamic communication protocol selection. In Training Set Classification Data Replication, a lot of multimedia data processing and number crunching is done through the pipelines of DMARF. This data is then pass through sample loading and classification stages where heavy CPU computation is done at preprocessing, feature extraction and classification stages. The classification stage absorb lot of data. In DMARF case this data may store on different hosts where there is chance of re-computation of data which is already computed by another classification host. In order to avoid this the classification stage node has to communicate to all classification members to exchange data. By this optimization of data mirroring/replication a lot of computational effort is saved at the end nodes. Dynamic Communication Protocol Selection also plays an important role in self-optimization by automatically selecting the most efficient available communication protocol in run time environment [2].

Distributed Modular Audio Recognition Framework (DMARF) has different levels of basic front-ends, from higher to lower so that a client application and services can invoke other services via front-ends while running in a pipeline-mode. It also has Back-ends that can provide actual servant implementations like primary-backup replication and disaster recovery etc. These distributed services can also intercommunicate.

In DMARF there is a disadvantage that if a pipeline stage process crashes out then the whole process will be unavailable or it may get lost forever, also information about the pending transaction will also be lost. To solve this problem Write Ahead Log (WAL) has been developed for DMARF which basically do backup replication so that information regarding uncommitted transaction can be restored using its replica manager.

As WAL is a prototype application therefore some assumptions has been taken that were not a part of its specifications [6]:

1. No garbage collection on server side in terms of limiting the WAL size.
2. No module other than Classification has WAL functionality.
3. No nested transaction in MARF services while pipelining.
4. No intercommunication between services other than pipeline mode.
5. No replication is present.

RMI (Glossary 2) and CORBA (Glossary 1) clients and servers use a command-line interface, GUI interface will be integrated in near future and its hardware interface is abstracted by JVM making DMARF architecture independent [6].

Synchronization is important for applications that allows access to a shared resource / data structure by multiple clients like the DMARF do so the synchronization is being maintained when database objects are being accessed from server by more than one client(s).

The tests which are being performed on DMARF has produced positive results and stopping any service replicas and restarting it has resumed the normal operation of pipeline in batch mode. DMARF is currently being implemented using three distributed technologies (RMI, CORBA and Web Services) but in future it will also be implemented for other communication technologies like JINI, JMS, DCOM+, .NET Remoting) and advanced features of distributed systems like fault tolerance, high availability will also be implemented. [7]

There are so many important and significant applications of DMARF like High-Volume processing of recorded audio, or imagery data is also possible, biometric applications are also present for DMARF. Using DMARF a bulk of recorded phone conversations in a police department can also be processed for subject identification, also portable devices like laptop, PDA and cellphones that does not have much high-performance can also be used to upload voice data on servers of DMARF network, which can be really helpful for investigators. [6]

### 3) *DMARF with ASSL*

The paper [6] is discussing about the autonomic behavior in DMARF by adding an autonomic middleware layer to DMARF pipeline using Autonomic System Specification Language (ASSL), the autonomic elements will be managed by a distinct autonomic manager.

The autonomic computing (AC) [3] is basically used to add features like self-regulation and to reduce workload that is required to maintain a complex system by changing them into self-managing autonomic systems. The ASSL [4] [6] is the tool by which these features can be implemented that is why it is being used in DMARF. The ASSL complete specification compiles it into wrapper Java code in order to fulfill the autonomic requirement by providing an autonomic layer to DMARF.

ASSL was used to develop autonomic features and generate prototype models for two NASA missions -the ANTS (Autonomous Nano-Technology Swarm) concept mission and the Voyager mission.

In both cases, there have been developed autonomic prototypes to simulate the autonomic properties of the space exploration missions and validate those properties through the simulated experimental results. The targeted autonomic properties were: self-configuring, self-healing, and self-scheduling for ANTS, and autonomic image processing for Voyager.

The autonomic system (AS) is basically composed of autonomic elements (AEs) that communicate over interaction protocols, hence the ASSL has defined it through formalization tiers. A multi-tier specification model is being provided by ASSL that is designed to be scalable and configures mechanisms and infrastructure elements needed by an AS.

The ASSL's multi-tires system architecture (AS) including formal language constructs to specify service-level objectives (SLO), core self-CHOP i.e. (self –configuration, self-healing, self-optimization and self-protection) autonomic properties, corresponding architecture, allowed actions, events and metrics to aid the self-management aspect of the system[8].

In order to add automicity to DMARF a special autonomic manager (AM) has been added to each DMARF stage making it ADMARF (i.e. autonomic DMARF). Self-healing is also been implemented by using ASSL, which provides following features:

• It monitors run-time performance and informs DMARF stages to start self-healing in case of performance degradation.
• If a node is down or not performing well then it will be replaced with a specified algorithm.

Although self-healing feature is being added to DMARF but it has not been completely implemented yet, after full implementation it will be able to fully functional in autonomous environments like on Internet, large multimedia processing firms, robotic spacecraft and for pattern-recognition systems so that these system can run for multiple days unattended.[6]

### 4) *GIPSY*

The General Intensional Programming System (GIPSY) is a framework for the compilation and distributed demand-driven evaluation of context-aware declarative programs [9]. GIPSY was designed and implemented mainly to achieve the goals of generality, efficiency and flexibility. GIPSY is a multi-language program environment executing multiple programming languages, like Lucid, C++ and Java programming language [10]. GIPSY was based on three sub systems that includes the General Intensional Programming Language Compiler (GIPC), General Eduction Engine (GEE) and the Intensional Run-time programming environment (RIPE).

### 5) *Architecture of GIPSY*

The compiler for GIPSY is known as GIPC and is based on GIPL (Generic Intensional Programming Language). The GIPC compiles the GIPSY program in two stages. In first part it translates the intensional part of the GIPSY in to C program and then the C program is further compiled in its default way.

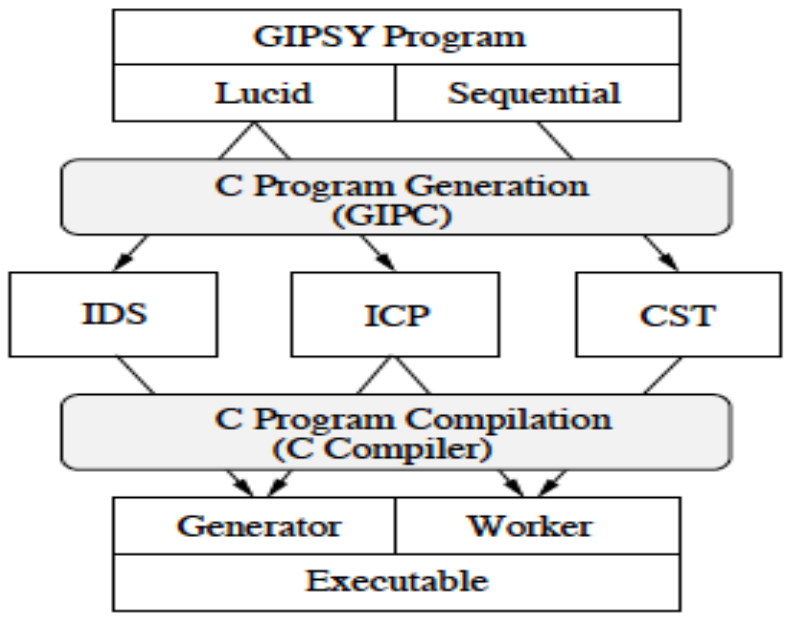

*Figure 2 Architecture of GIPSY describing compilation process*

In **Fig. 2,** the program consists of two parts: the Lucid part and the Sequential part. Lucid part is compiled into intensional data dependency structure (IDS), which describes the dependencies between the variables whereas the Sequential part defines the granular sequential computational units. Data communication procedures used in a distributed evaluation of the program are also generated by the GIPC according to the data structures definitions written in the Lucid part, yielding a set of intensional communication procedures (ICP) [11]. GIPSY was based on demand-driven model of computation in which the computation takes place only if there is a demand for it. To overcome this overhead GIPSY uses eduction engine that in conjunction with demand-driven computation uses a value cache that stores the computed values in it. In GEE each demands generates a procedure call that reduces the overhead [11]. The distributed runtime system of GIPSY includes a Demand Migration Framework (DMF) also known as Demand Migration System (DMS) that has been presented using the two technologies, namely JINI and JMS [9] [20].

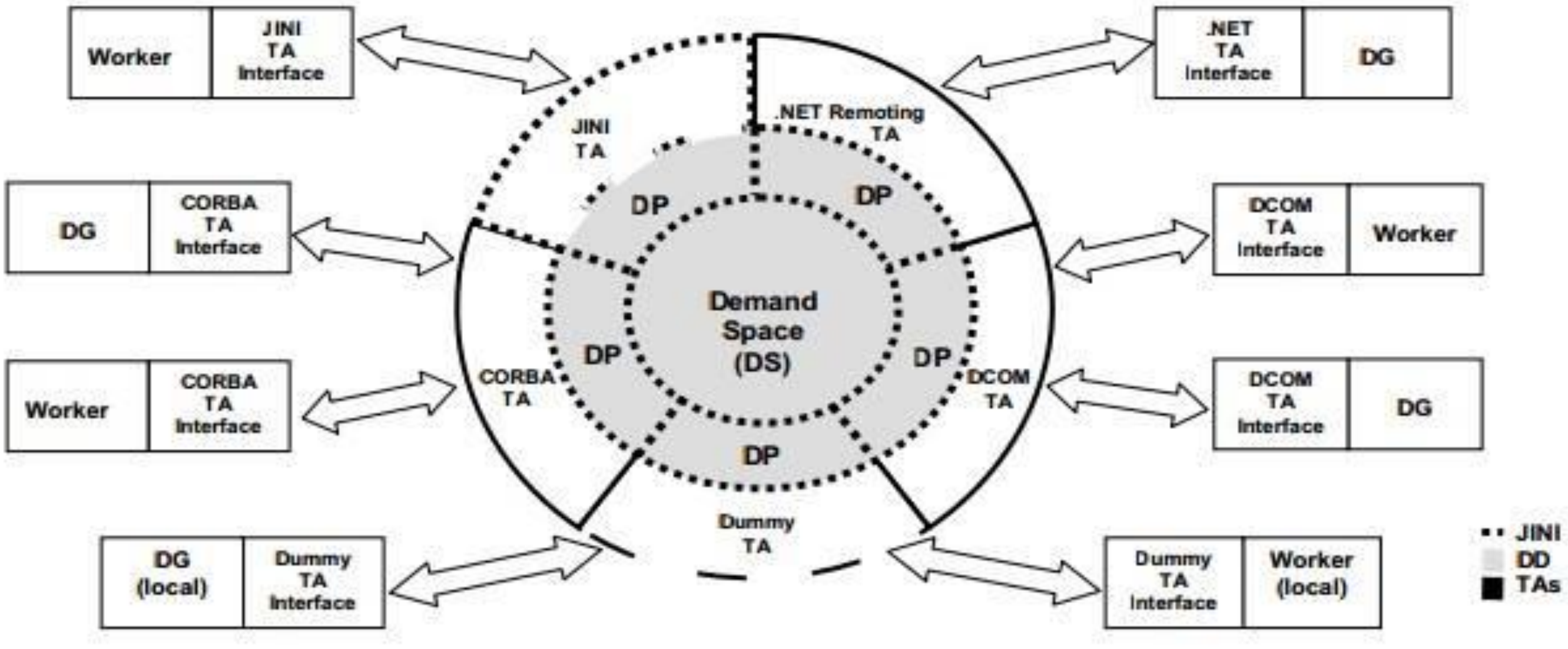

*Figure 3 Architecture of DMF (Demand Migration Framework)*

JINI is a network oriented computing technology used for the construction of distributed systems. JINI forms "an open architecture for handling resource components within a network" [13]. As JINI and GIPSY are developed in Java, they get integrated easily. JINI depends on JINI Lookup Service (LUS) for managing and handling various services [12] [13]. JMS is a Java based Message oriented middleware technology [14] [15] [16] JMS broker is JMS implementation that provides messaging service to the clients. Similar to JINI, JMS also supports persistence. In **Fig3.[20]**, the large circle represents the DMS and the rectangles on both sides of the circle form the DG (Demand Generators) and workers. The DMS is formed of two independent contributors namely DD (Demand Dispatcher) and TA (Transport Agent). In the Fig.2 the area shaded with the gray color represents the DD. This DD uses the TA to deliver the messages. They together form a communication system for the DMS. TA`s are not required when the Generators and Workers communicate in local.

The autonomic computing is essential in the current technology era. GIPSY can also be made autonomic. Making GISPY autonomic is the main goal of GIPSY in which it makes GIPSY to be self-learning and self-manageable [17]. One of the core features of Autonomic Computing (AC) is self-monitoring. Autonomic GIPSY allows an automated monitoring and management of the complex multi-tiered, distributed, heterogeneous workloads across the GIPSY infrastructure to better achieve assigned processing performance goals for end-user services. GIPSY is a multi-tiered architecture. There are many different task and tiers in GIPSY. Each tier shares a separate task at the time of execution. The tiers in GIPSY spawn processed, these processes communicate with each other through demands. There are four different tasks in GIPSY which are illustrated in the figure below:

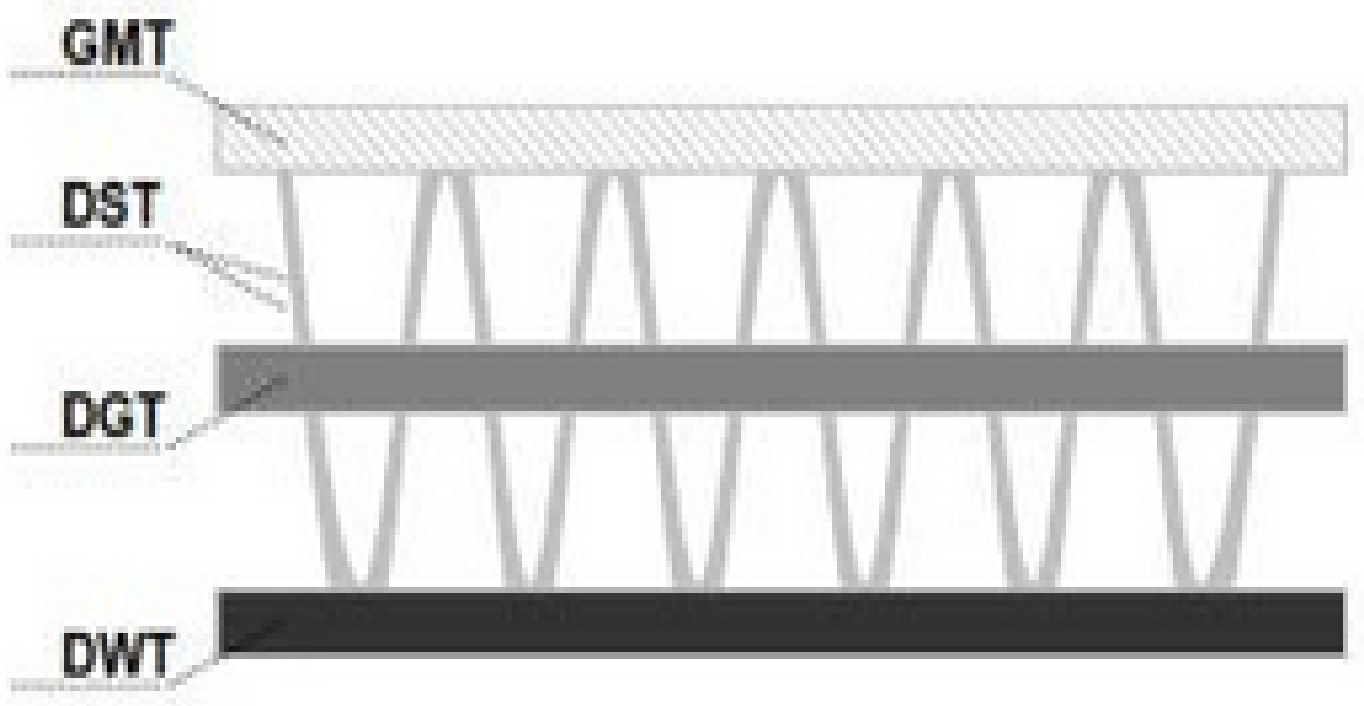

*Figure 4 Multi-tier Architecture of GIPSY*

**GIPSY Manager Tier (GMT)** - GIPSY managers are instances of this tier which enable registration of a GIPSY node to a GIPSY instance and allocation of various GIPSY tiers to these nodes.

**Demand Store Tier (DST)** - It is a middle ware tier that exposes other tiers to GIPSY Demand Migration System (DMS). This is a communication system that connects tiers through demands.

**Demand Generator Tier (DGT)** – Instances of this tier encapsulate in itself a demand generator. The purpose of this is to generate demands.

**Demand Worker Tier (DWT)** – Instances of this tier contain a set of demand workers that process procedural demands resulting in a compiled version.

*6) ASSL with GIPSY*

To automatically augment the ASSL (Autonomic System Specification Language) framework self-forensics as an autonomic property is used for formal specification tools for autonomic system [8]. The idea and implementation of self-forensic with ASSL and GIPSY (General Intensional Programming System) is currently at the conceptual level. The ASSL framework takes input in the form of specification of properties of autonomic system after that the ASSL checks the specifications syntax and semantics formally and if the specifications are ok then according to the corresponding specifications Java collection of classes and interference are generated.

Now comes the part of Self-Forensics concept which has the forensic logging of Forensic Lucid. Forensic Lucid was initially used for specification, automatic deduction and event reconstruction in the cybercrime domain of digital forensic [8].

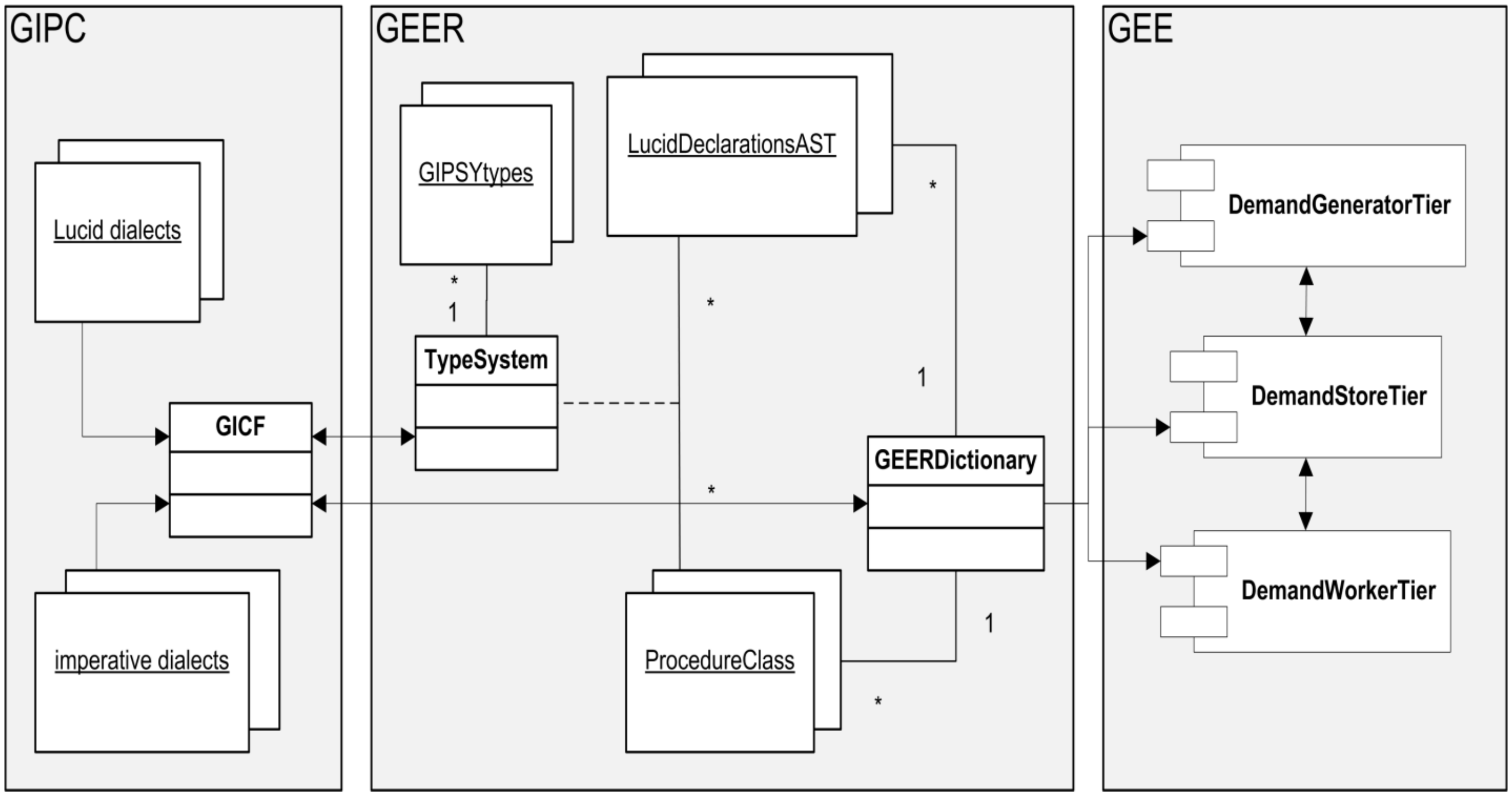

*Figure 5 GIPSY High Level Overview*

Now comes the part of Self-Forensic autonomic property but in order to use it we need to add self-forensic policy specification for AS tier and also for AE tier. In order to translate forensic events we first have to add syntax as well as semantic support for lexical analyzer, parse and semantic checker of ASSL also we have to add code generator for JOOIP and Forensic Lucid.

## B. Summary

Metrics are the measurements applied to the DMARF and GIPSY source codes. The tools used for the measurement is defined in Table 1.

Table 2, shows the methodologies which we used to measure both DMARF and GIPSY projects source code. The metrics calculated in this document are Number of Java Files, Number of Classes, Number of Methods, Number of Java lines of code.

We have used "Kalimetrix Logiscope" for determining the metrics. Logiscope is a metrics tool used to calculate C, C++, Java and Ada Code. It provides metrics in the number, graphical formats. In major, it consists of four products namely Quality Checker, Rule Checker, Test Checker and Code Reducer. Quality Checker anticipates and identifies the problems by providing information based on the metrics and graphs. Using the Perl and Tcl languages Rule Checker define and verifies the coding rules. The Test Checker improves the code reliability by specifying the untested code. And finally, the Code Reducer filters the code and identifies the duplicate code. It does re-factoring and save maintenance costs.

| Metrics | Methodology |
|---|---|
| Number of Java Files | Logiscope |
| Number of classes | Logiscope |
| Number of Methods | Logiscope |
| Number of Java lines of code | Logiscope |

*Table 1 Tools used for the Measurement*

| Metrics | DMARF | GIPSY |
|---|---|---|
| Number of Java Files | 201 | 602 |
| Number of classes | 319 | 956 |
| Number of Methods | 2419 | 6448 |
| Number of Java lines of code * | 24479 | 104083 |

*Table 2 Measurements or Metrics for DMARF and GIPSY*

Due to the number of classes and methods, GIPSY project is larger than DMARF project. All measurements, Number of classes, Number of Methods, and lines of code for GIPSY project are almost three times greater than DMARF project.

1) *B.A Team Work and References*

| Team Name | GIPSY Reference | DMARF Reference |
|---|---|---|
| Osama Yawar | [11] | [8] |
| Tahir Ayoub | [9] | [2] |
| Parham Darbandi | [21] | [18] |
| Yashwanth Beeravelli | [20] | [19] |
| Nadir Muhammad | [22] | [7] |
| Shahroze Jamil | [23] | [6] |

*Table 3 Individual Readings of GIPSY and DMARF*

The above mentioned references of GIPSY and DMARF were read by us and integrated all the summaries into the backgrounds mentioned in Section II.

## III. CONCLUSIONS

In conclusion, GIPSY provides many features such as scalability, flexibility, usability and extensibility. In addition, GIPSY is a multi-language supporting environment. It executes Java, C++ and Lucid. It can be made autonomic for self-monitoring. GIPSY also supports demand driven execution in distributed and heterogeneous environments. In a pure demand-driven model of GIPSY there were several overheads which were reduced by using eduction in conjunction with demand-driven computation.

Also, DMARF provides many features like self-regulation to reduce workload as it has to process huge amount of voice data. It is a significant and useful framework that can be used in portable devices as well that has low computational power. Despite all that the research is still going on DMARF and it will be implemented for other communication technologies as well and DMARF will also be fully functional in autonomous environments.

## IV. REQUIREMENTS AND DESIGN SPECIFICATIONS

### A. Personas, Actors and Stakeholders

1) *DMARF Personas, Actors and Stakeholders*

**Persona:** A Persona is a fictional character played by an actor. The concept of Persona is used in Software Requirements Elicitation. It is considered as one of the best techniques to model a user.

a) ***DMARF Persona***

| **John (Audio Specialist Engineer)** | |
|---|---|
| 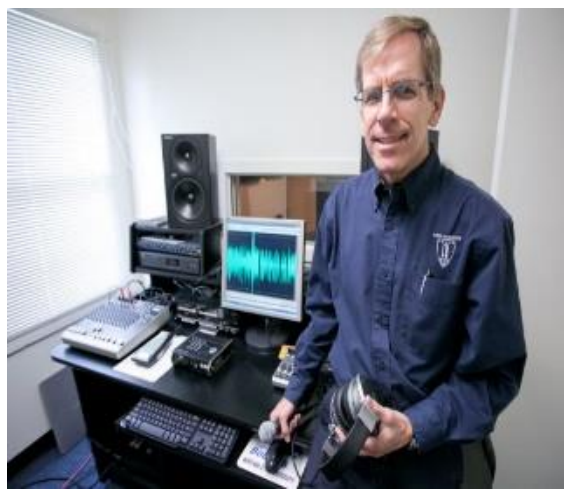 | **Age**: 40<br>**Qualification**:<br>Masters of audiology<br>**Company**:<br>United States Secret Service<br>**Technology Level**:<br>Very familiar and comfortable with the technology |
| **Description**<br>John is an audio engineer at US Secret Service. He is very flexible with the technology. He is also doing his PhD at Oklahoma State University. He is doing his research work on audio recognition tools and would like to use it at his work.<br><br>The audio should recognize a specific voice from a set of voices provided as input to it. It should work accurately even the user provides female and male voice.<br><br>His main task is to identify the person whose voice has been captured in order to help his agency to find out criminals by matching the captured voice with the voices that are available in records of United states secret service. | **Highlights**<br>Doing PhD at Oklahoma State University Also working as Audio Specialist Engineer at US Secret Service<br>Develops new technologies for the Secret Service Helps his supervisor in research work<br><br>**Goals**<br>Handle the cases using his audio recognition tools before deadline Complete the assignments<br>Complete his PhD by 2016 Complete his thesis on audiology.<br>Develop audio recognition tool for the Secret Service |

*Table 4 DMARF Persona*

### b) *DMARF Actors:*

- **DMARF based system**: DMARF capture autonomic functioning of the distribute pattern recognition pipeline and its optimization.
- **DMARF Developer**: DMARF Developer try to develop middleware, compiler, new algorithms, and framework. They can be external people of the outside of the organization.
- **Application Developer DMARF**: Application Developer DMARF are the second level developers who are using DMARF framework. Code venerable detection, speaker identification
- **Professor***: Professors are supervising the projects as well as funding the project. However, they themselves are not necessary developing, they may be participating in design or architecture review.
- **Student***: Student can be a DMARF developer or DMARF application developer, but developers might not be necessary students.
- **Researcher***: Researchers can be academic within a university or outside. They do test algorithms, test compilers, or use middleware as a demand driven application.

*. Some of them can be developers.
It is possible there is an overlapping among the actors.

### c) *DMARF Stakeholders*

- **Investigator**: An investigator has the ability of uploading from, e.g., a laptop, PDA, or cell phone collected voice samples to the servers constituting a DMARF-implementing network.
- **Project Manager**:
  All organizations and people are funding the project.

## 2) *GIPSY Personas, Actors and Stakeholders*

### a) *GIPSY Persona*

<table>
<tr><td>Peter(Researcher)<br>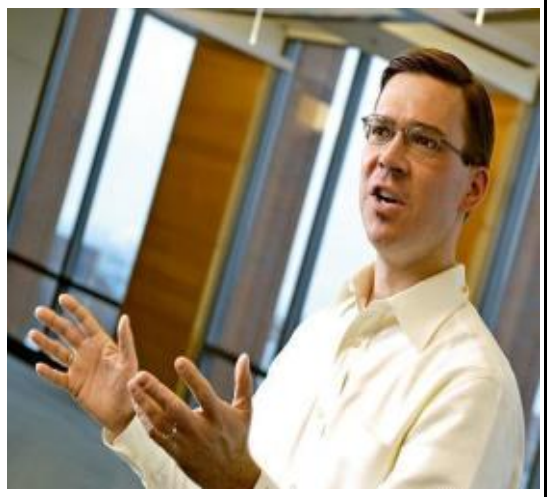</td><td>Age: 32<br>Qualification: Masters of Information Security<br>Organization: University of Toronto<br>Level of Technology: Very comfortable</td></tr>
<tr><td rowspan="2">Description<br>Peter is a PhD student at University of Toronto. He is doing his research work on Programming Intensional Logic and intent to submit his thesis by March 2017. He is interested in learning new, currently introduced, technologies and programming languages. Currently he is working on a framework which supports multiple languages, multitier architecture and demand driven architecture. This demand driven architecture should provide demands and the other components working on the demands. His study involves modular collection of frameworks that provides sustainable support for intensional programming languages based on multi-tier architecture. His research also involves how intensional programming languages can adopt to a distributed environment and their support to it.</td><td>Highlights<br>Doing PhD on Intensional Programming at University of Toronto Assisting the professor in extending his research.<br>Highly interested in learning new technologies and using them. Working as a Lecturer as well as Teaching assistant.</td></tr>
<tr><td>Goals<br>Creating a Intensional logic System<br>Submit the thesis by March 2017<br>Get feedbacks from students<br>Teach in a precise manner. Work on the assignments assigned by the supervisor and submit them before deadline</td></tr>
</table>

*Table 5 GIPSY Persona*

*b) GIPSY Actors:*

- **GIPSY Developer**: GIPSY Developer try to develop tiers, middleware, compiler, new algorithms, and framework. They can be external to the organization
- **GIPSY Application Developer**: GIPSY Application Developer is the second level developer who use GIPSY framework and evaluate heterogeneous programs
- **Supervisor***: Supervisors supervise the project as well as sponsor the project. However, they themselves are not necessary developing, they may be participating in design or architecture of GIPSY
- **Student***: Student supports the Supervisor and GIPSY developer in developing the framework
- **Researcher***: Researcher can be academic within a university or outside. She/he do test algorithms, test compilers, or use middleware as a demand driven application. They may be sponsor sometimes

***Some of them can be developers.
***It is possible there is an overlapping among the actors.

*c) GIPSY Stakeholders:*

All human actors and
- GIPSY Developer
- Supervisor
  - Professor
  - Architect
  - Designer
  - Tester
- Researcher
  - Project Manager
  - Sponsor
- GIPSY Application Developer
  - Software Developer
  - Student

## B. Fully Dressed Use Cases

### 1) DMARF Fully Dressed Use Case:

| | |
|---|---|
| **Use Case ID** | UC – DMARF – 01 |
| **Use Case Name** | Training set classification data replication. |
| **Scope** | The system under design |
| **Level** | User level |
| **Primary Actor(s)** | Application Developer DMARF |
| **Supporting Actor(s)** | **DMARF based system**: DMARF capture autonomic functioning of the distribute pattern recognition pipeline and its optimization.<br>**Autonomic researcher**: They do test algorithms, test compilers, or use middleware as a demand driven application<br>**Professor**: Professors are supervising the projects as well as funding the project. |
| **Stakeholder and Interests** | **Investigator**: An investigator has the ability of uploading from, e.g., a laptop, PDA, or cell phone collected voice samples to the servers constituting a DMARF-implementing network.<br>**SpeakerIdentApp** application is Text-Independent Speaker Identification (who is the speaker, their gender, accent, spoken language, etc.) |
| **Success Guarantee** | Classification stage nodes communicate to exchange the data among all the members. |
| **Brief Description** | Avoid of heavy computations in the pre-processing, feature extraction, and classification stages which do a lot of CPU-bound number crunching and matrix operations. |
| **Pre-Conditions** | |
| Audio/voice sample(s) to the servers constituting a DMARF-implementing network. | |
| **Post-Conditions** | |
| **Success End Condition**<br>DMARF can use data already computed data on another classification host. | |
| **Failure End Condition**<br>DMARF recomputed data which is already computed on another **classification host**. | |
| **Main Success Scenario** | |
| 1. User indicates to the system that he/she wants forensic analysis and biometric subject identification.<br>2. System notifies user that he/she can upload file.<br>3. User uploads the data file.<br>4. System indicates that data file has been uploaded successfully.<br>5. DMARF classification stage node communicates to different hosts and checks for same but already computed data.<br>6. DMARF classification node finds the data and starts acquiring it from different hosts. | |
| **Alternate Scenario** | |
| 3.a. User choose to cancel the file upload.<br>3.b. System notifies the user that the user file upload request has been cancelled.<br><br>4.a. System notifies the user the file uploaded is corrupt.<br>4.b. Return to step 3.<br><br>6.a. DMARF classification node did not find the data on any host.<br>6.b. DMARF computes the data. | |
| **Special Requirements** | Audio/textual or imagery data samples.<br>Classification hosts. |
| **Technology and Data Variations List** | Desktop computer, PDA, Cell phone, Laptop, RMI, CORBA, Java, Audio (Mp3, wmv, wav, etc.)/textual/imagery files. |
| **Frequency of Occurrence** | High |
| **Miscellaneous** | Work is still going on DMARF and ASSL so that more realization of ADMARF will be done. |

*Table 6 DMARF Fully-dressed Use Case*

2) ***GIPSY Fully Dressed Use Case***

| | |
|---|---|
| **Use Case ID** | UC – GIPSY – 02 |
| **Use Case Name** | Generates **Demand** |
| **Scope** | General Intensional Programming System |
| **Level** | System Level |
| **Primary Actor** | **GEE Generator** |
| **Secondary Actor** | **GEE Worker** |
| **Stakeholders and Interests** | GIPSY Developer, Researcher |
| **Success Guarantee** | **Demands** are properly generated and propagated. |
| **Brief Description** | |
| GIPC (General Intensional Programming Compiler) generates IDS (Intensional Data Dependency Structure) that is interpreted by GEE (General Eduction Engine) at run-time to generate the demands. The demands are then propagated by IDP (Intensional Demand Propagator) and are sent to the worker. | |
| **Pre-Conditions** | |
| GIPC generates IDS | |
| **Post-Conditions** | |
| **Success End Condition**<br>IDP generates, propagates **demands** and send it to worker using ICP (Intensional Communication Procedures). | |
| **Failure End Condition**<br>**Generator** and **Worker** are not platform independent and flexible with distributed computation technologies, **demands** are not interpreted | |
| **Main Success Scenario** | |
| 1. User indicates to the system that he/she wants to provide **GIPSY program code** as input to the GIPSY compiler<br>2. **GIPC (GIPSY compiler)** converts the **Lucid part** of the code into IDS<br>3. GEE interprets IDS<br>4. The GEE generator evaluates the low-charge RIPE sequential threads locally<br>5. The **demands** are generated by the **GEE generator**<br>6. The IDP propagates demands and sends it to **GEE** worker using ICP | |
| 7. **Demand** is placed in a queue and removed after its computed successfully | |
| **Alternate Scenario** | |
| 1.a.1 GIPSY compiler throws an exception<br><br>1.a.2 User provides the input<br><br>3.a.1 GEE does not interpret IDS<br><br>4.a.1 **GEE generator** will not be able to evaluate<br><br>4.a.2 Return to Step 3<br><br>7.a.1 Demand remains in the queue to be computed<br><br>7.a.2 Demand is prioritized and removed following calculated | |
| **Technology and Data Variations List** | Lucid, C, Java |
| **Special Requirements** | GIPC (GIPSY compiler) |
| **Frequency of Occurrence** | High |

*Table 7 GIPSY fully-dresses Use Case*

**Actor:**

A Supervisor of GIPSY may be a Professor, scientist. He/she is involved directly in the project. A Supervisor develops the architecture of the project, design the project, responsible for the whole core architecture of the GIPSY. From the time of proposal, Supervisor is responsible till the testing and release. The GIPSY project may or may not be validated by the experts. A Supervisor is responsible to describe the GIPSY, its uses, drawbacks, etc.

## C. Domain Model

### 1) *DMARF Domain Model Description:*

**Sample:** could be a high-volume processing of recorded audio, textual, or imagery data.
The SampleLoadingNode accepts incoming voice, text, image, or any binary samples and does process on it and stores the result in SampleLoadingHost and then preprocessingNode accepts data from SampleLoading and does the requested preprocessing (a variety of filters, normalization, etc.).
The Feature Extraction Service accepts data, presumably preprocessed, and attempts to extract features out of it given requested algorithm (out of currently implemented, like FFT, LPC, MinMax, etc.) and may optionally query the preprocessed data from the Preprocessing Service.
The Classification and Training Service accepts feature vectors and either updates its database of training sets or performs classification against existing training sets. It may optionally query the Feature Extraction Service for the feature vectors.
**Computing devices**: could be a desktop computer or a mobile equipment such as a laptop, PDA, cellphone, etc. The Sample Loading Service knows how to load certain file or stream types (e.g. WAVE) and convert them accordingly for further preprocessing.
**Classification Host**: different hosts that run the classification service.

#### a) *DMARF Domain Model*

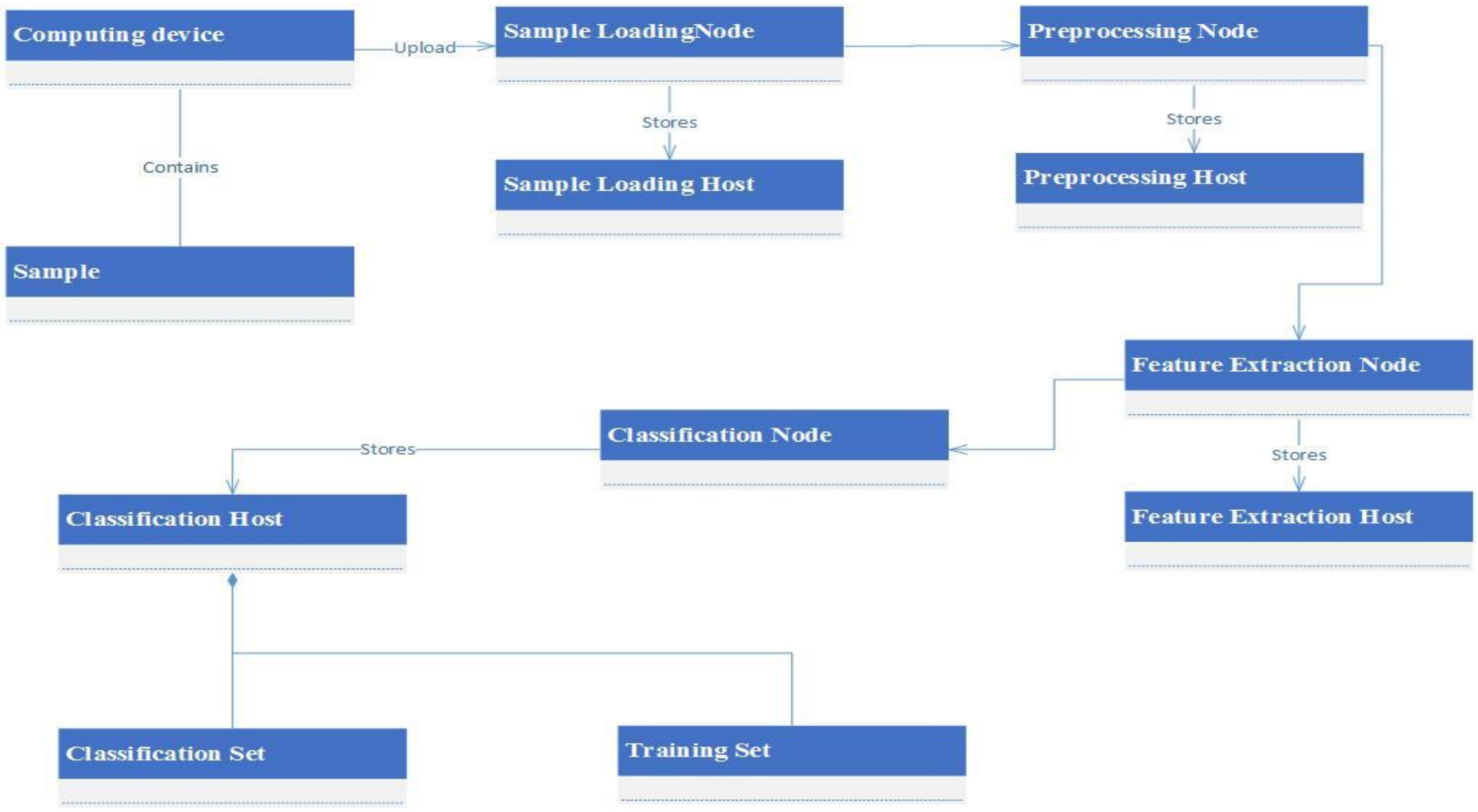

*Figure 6 DMARF Domain Model*

### 2) *GIPSY Domain Model Description:*

GIPSY is a framework for the compilation of heterogeneous intensional programs. The intensional programs may include Lucid, JLucid, Forensic Lucid, etc. These languages are compiled by the compiler of GIPSY i.e., GIPC (GIPSY Intensional Program Compiler). GIPC compiles the Lucid code and produces a .gipsy file similar to a .class file in Java. This .gipsy file is interpreted by GEE (General Eduction Engine). The .gipsy file is called GEER (General Eduction Engine Resource), as it is the resource for the GEE. This GEE Resource is represented by the GIPSYProgram in the design model

#### a) *GIPSY Domain Model*

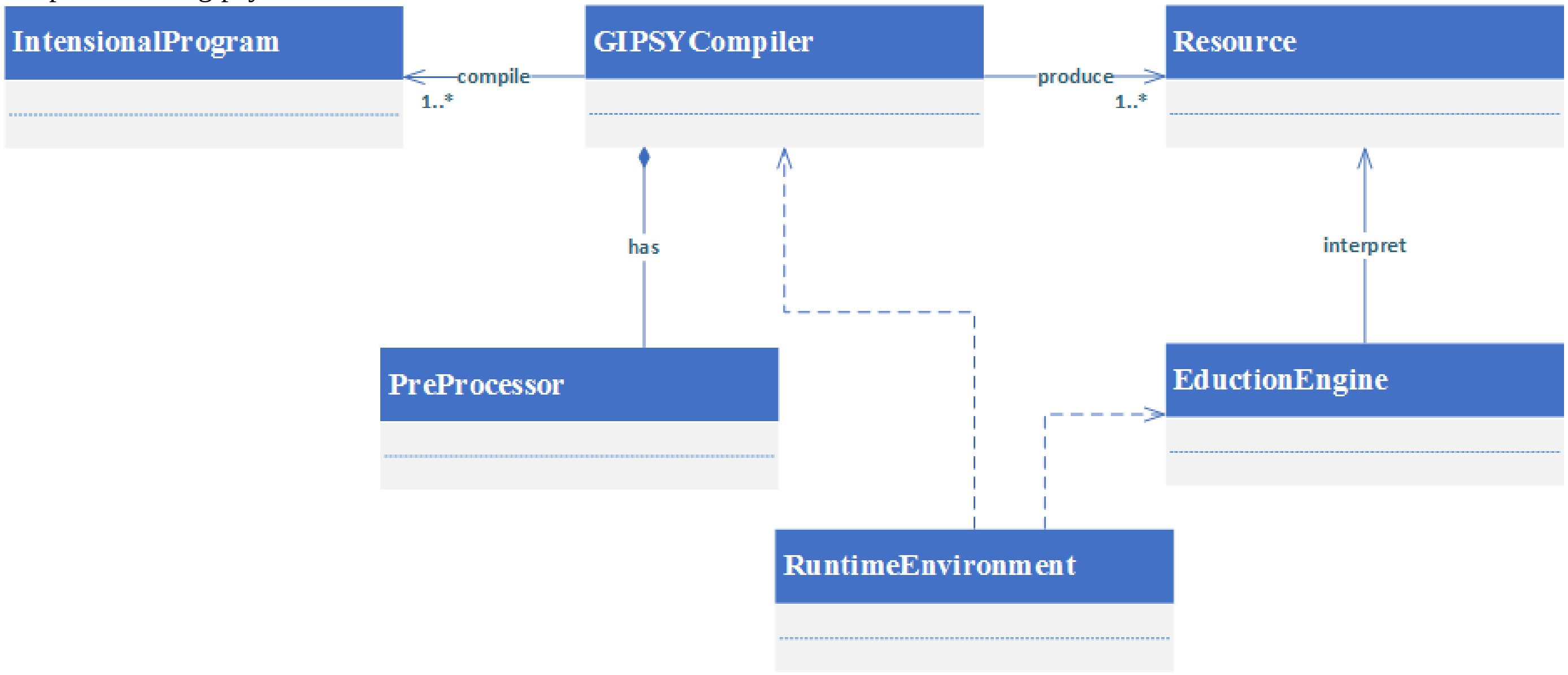

*Figure 7 GIPSY Domain Model*

### 3) *DMARF over GIPSY run-time architecture:*

#### a) *Description of DMARF over GIPSY:*

**Sample:** could be a high-volume processing of recorded audio, textual, or imagery data.
User is the one who generate such samples, which shall generate demand for SampleLoading. Sample is uploaded to DMARF computing device. Computing device **worker** checks for the result of same computing data in ClassificationDST (warehouse) if it founds there the result shall be shown to computing device which the user can view. In case the worker didn't find the result in ComputingDevice DST its DemandGenerator (DGT) shall generate the demand which is stored in warehouse (DST). In figure [8] each node of DMARF is split into Demand worker and Demand generator tier. The sampleLoaderNode's demand worker tier shall observe for the **pending** demands stored in DST by DemandGenerator of computing device and shall start working on it. The SampleLoading DWT shall store the result in SampleLoader DST. The SampleLoading DGT shall generate demand for PreprocessNode and store in SampleLoader DST which Preprocess DWT shall observe and shall start working on it and shall store the result in PreprocessingDST. After that the PreprocessNode DGT shall generate the demand for FeaureExtraction and shall store demand in PreprocessDST from where the FeatureExtraction DWT shall observe the work and start working on it and shall store result in FeatureExtraction DST. After that FeatureExtraction DGT shall generate the demand for ClassificationNode and shall store the demand in FeatureExtraction DST from where the Classification DWT shall observe the demand and shall start working on it and shall store the result in Classification DST. In Classification case we don't have Classification Demand generator but classification has ClassificationSet and TrainingSet. We store the TrainingSet in Classification DST.

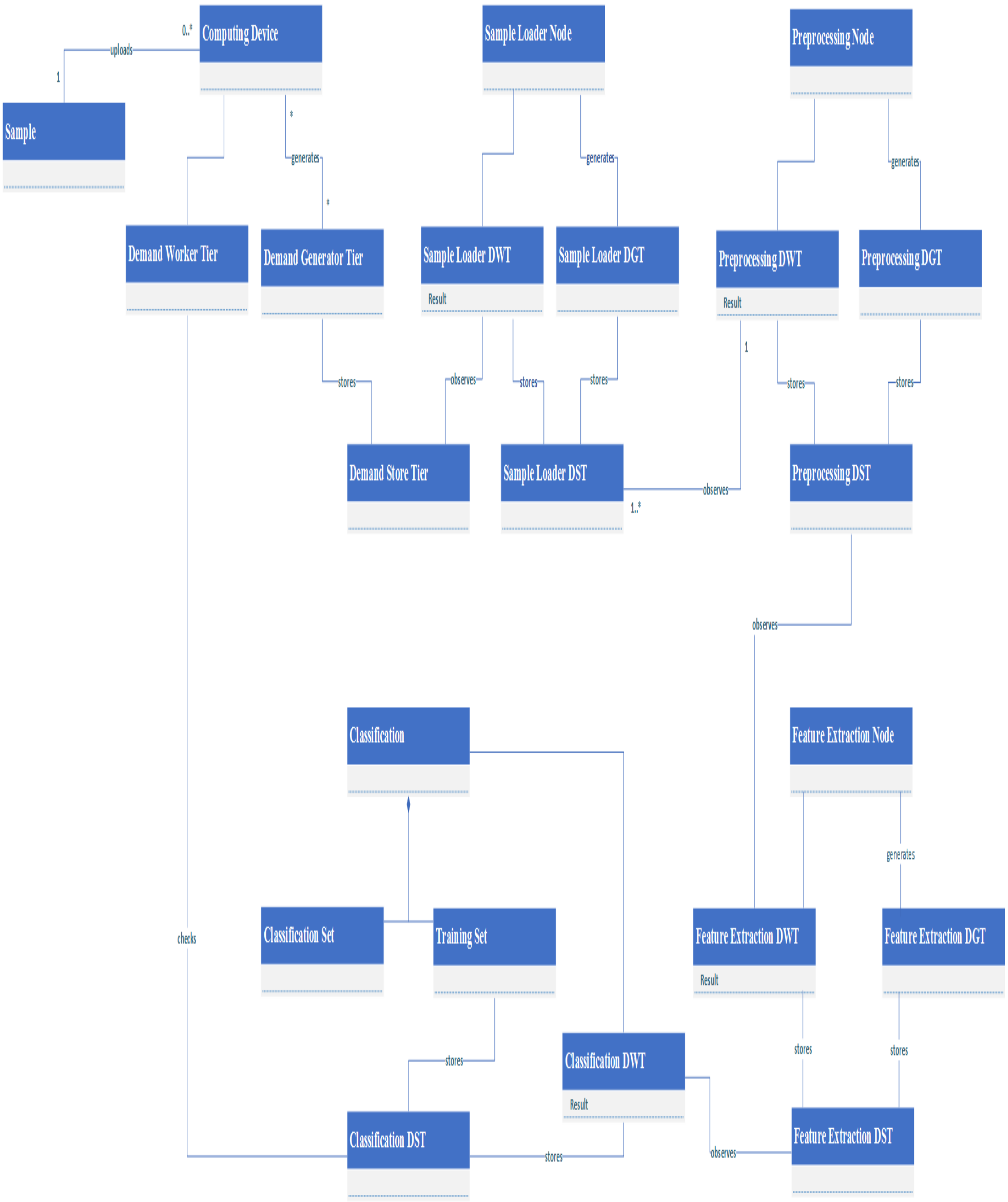

*Figure 8 DMARF over GIPSY run-time architecture*

**Differences between the concepts and the actual classes – GIPSY**

After doing comparison of domain model and class diagram of DMARF we found that there are no bigger differences in both conceptual and actual models as there are separate classes and packages for each operation of DMARF pipeline like SampleLoading, Preprocessing, FeatureExtraction and Classification approximately same as the domain model.

## *D.* Class Diagram

### *1) DMARF Class Diagram*

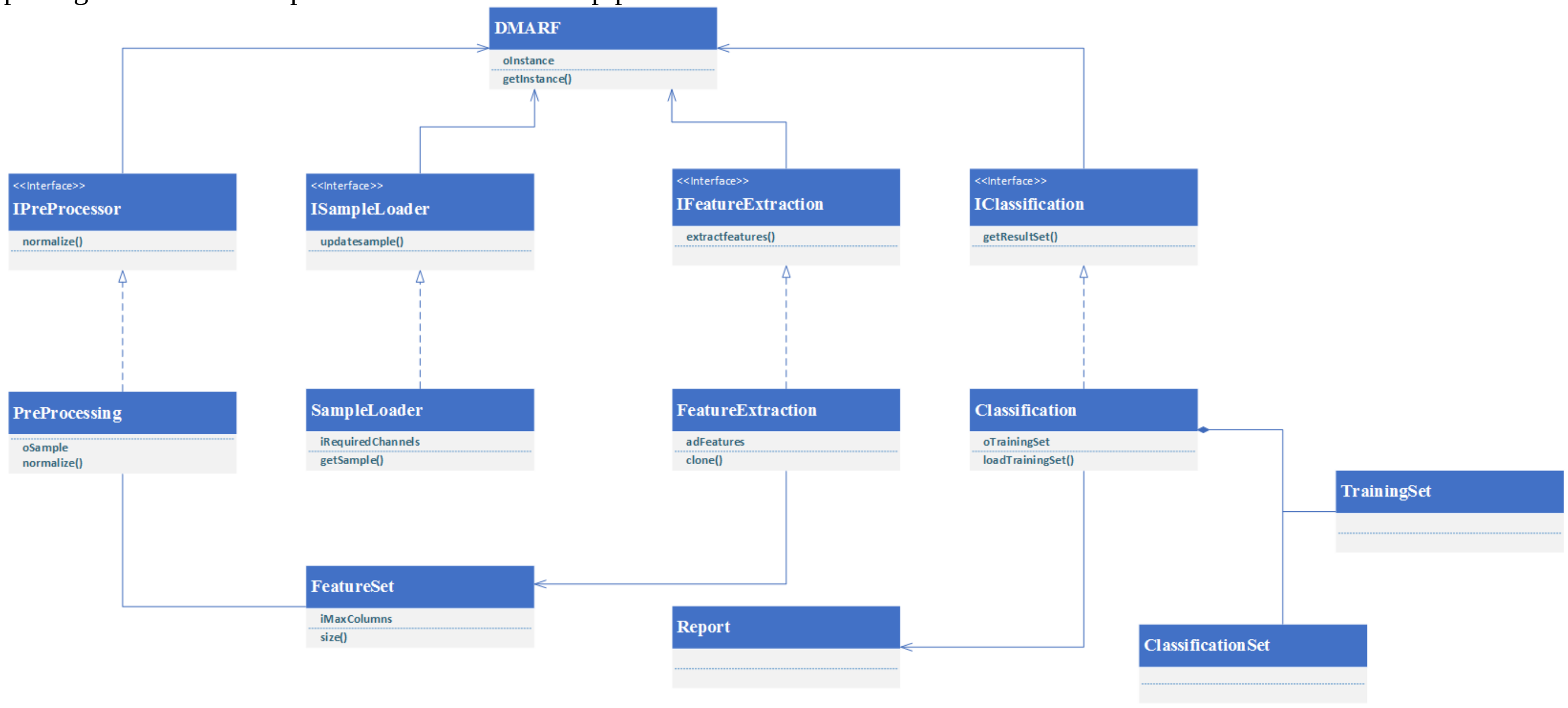

*Figure 9 DMARF Class Diagram*

**Description**

In DMARF Class Diagram the user has samples, samples can be in audio form. The user uploads the sample file for forensic analysis on computing devices. Computing device has MARF which has many interface namely:

- IPreProcessor
- ISampleLoader
- IFeatureExtression
- IClassification

Each interface has generalized feature like preprocessor, sampleloader, featureExtression and classification respectively. PreProcessor extract FeatureExtrection has feature sets which are extracted by the preprocessor class. Which classify the classification class. Classification class connects to different host to check the report for the sample uploaded by the user. If it found the report in storage it extracts the report from that host and show the report to user without doing any computation else it will compute it and makes the report.

### *2) GIPSY Class Diagram*

#### *a) **Gipsy Main Class Diagram (Built using tool objectAid)***

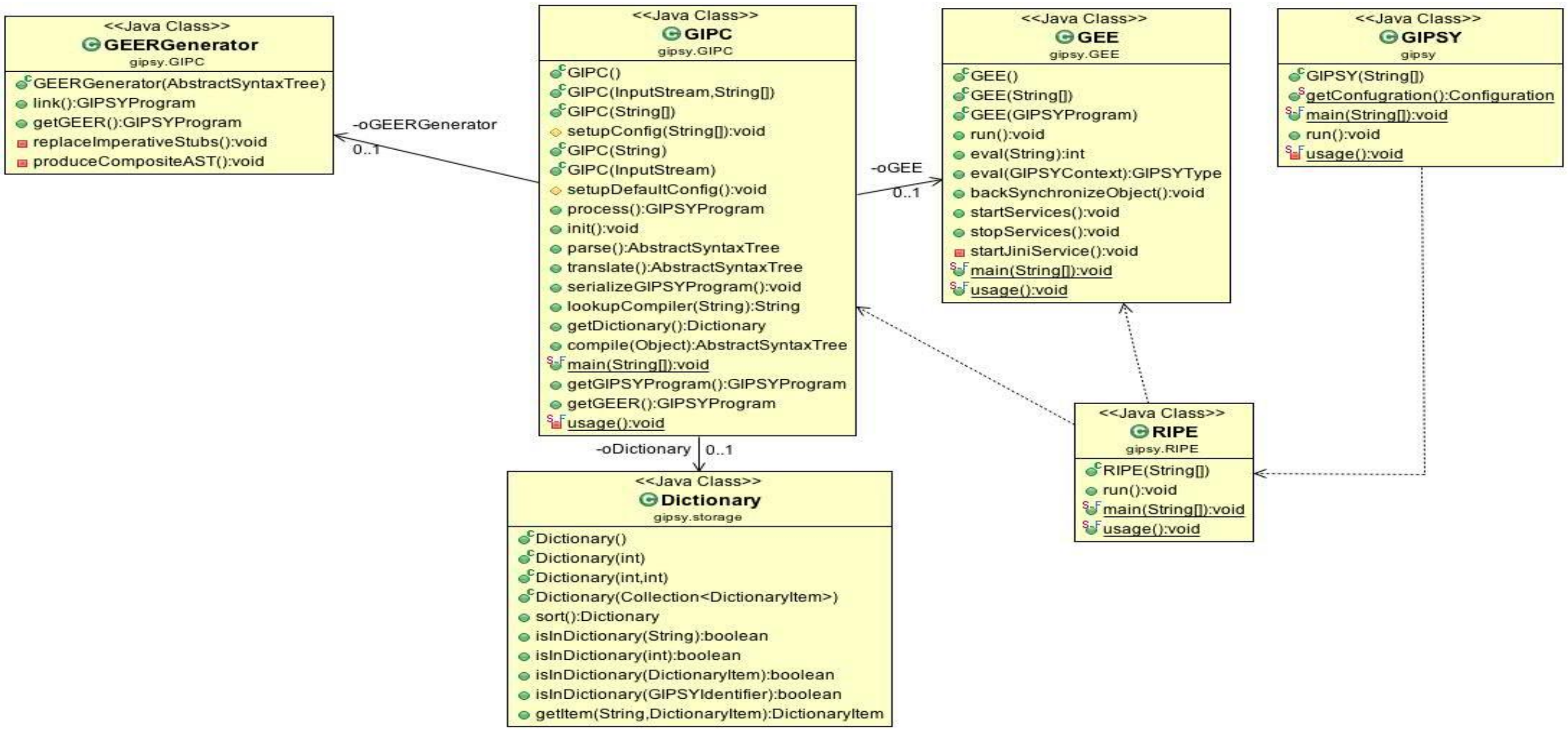

*Figure 10 GIPSY Main Class Diagram*

**Description (GIPSY Class Diagram):**

As discussed in the domain model, GIPSY consists of three major modules. They are GIPC, GEE, RIPE. GIPSY follows a modular approach to have the dynamic replacement of each and every component. The replacement takes place at compile time or run time leading improvement in the overall efficiency of the system.

In general, the Lucid code is accepted by the GIPC compiler and the specific compiler compiles the specific type of lucid. For example, ForensicLucid program is compiled by ForensicLucisd Compiler. All these compilers are available in the GIPC Class diagram. On the whole, GIPC compiles the code in a two stage process. The GIPSY intensional part is translated into C and this C program is compiled in the standard compilation method. Of the two parts in the source code the Lucid part defines data dependencies between variables and the sequential part that defines the granular sequential computation units. This Lucid part is compiled into a dependency structure due to the presence of dependencies. This dependency structure is named as Intensional Dependency Structure (IDS). It describes the dependencies between each variable participating in the Lucid part. GEE (General Eduction Engine) interprets this dependency structure. The second part of the GIPSY program are translated into C code using the second stage C compiler syntax resulting CST (C Sequential Threads). The modular design of GIPC allows sequential threads to be the programs in different languages. GIPC releases resource(s) for the GEE. This resource is GEER (General Eduction Engine Resource). The GEER consists of the .gipsy file or the output of the GIPC. It is similar to the .class file in a Java compilation process. RIPE (Run-Time Interactive Programming Environment) is a visual runtime programming environment. It enables the dataflow diagram related to the Lucid part of the GIPSY program. The GEER Generator creates the resources from the GIPC. GIPSYProgram represents the GEER. This GIPSYProgram is a container over the Abstract Syntax tree and Dictionary of the imperative nodes.

*b) **GIPSY Class Diagram (Built reading the case studies)***

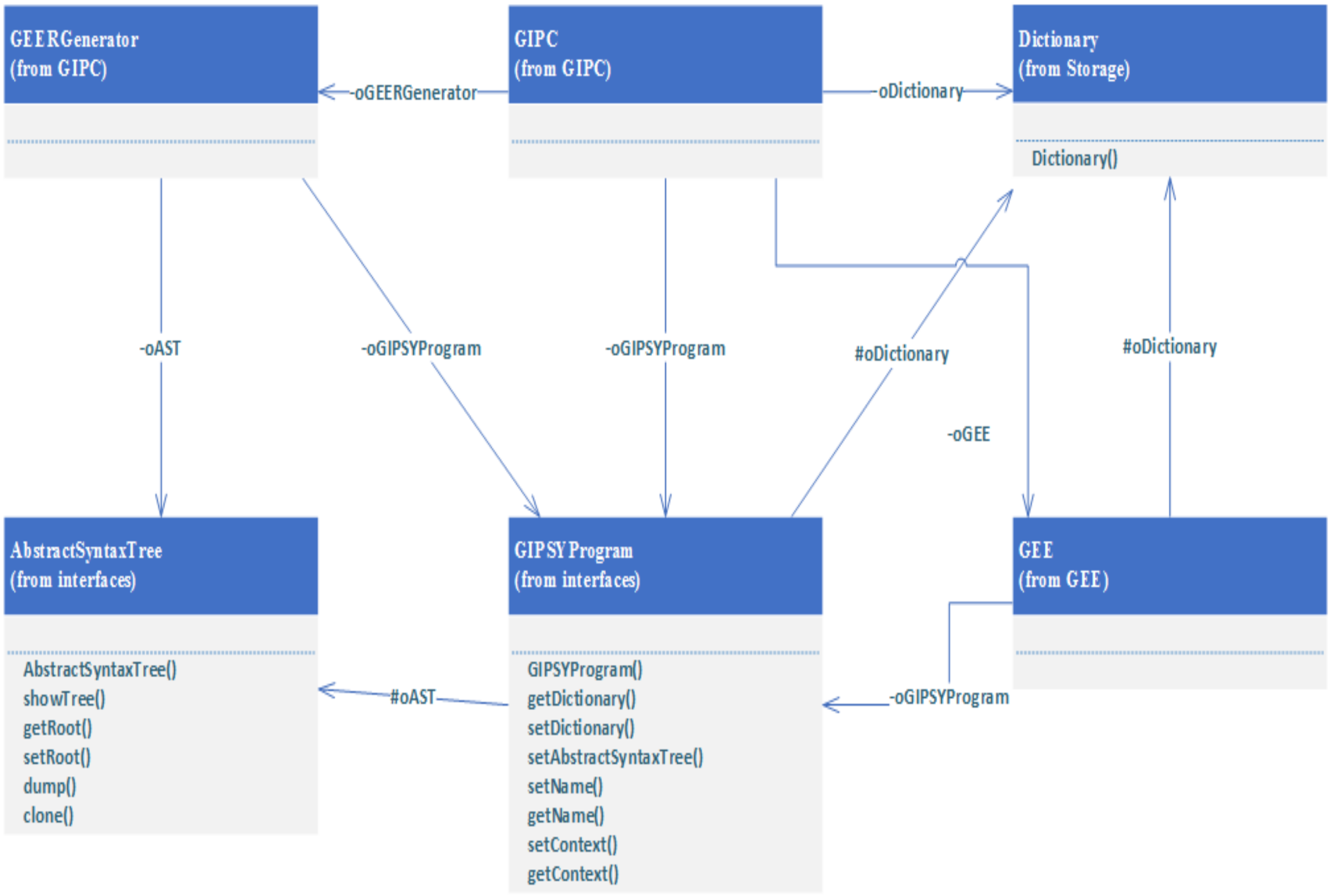

*Figure 11 GIPSY Class Diagram build by reading case studies*

### c) *GIPSY Compiler Class Diagram (built reading case studies)*

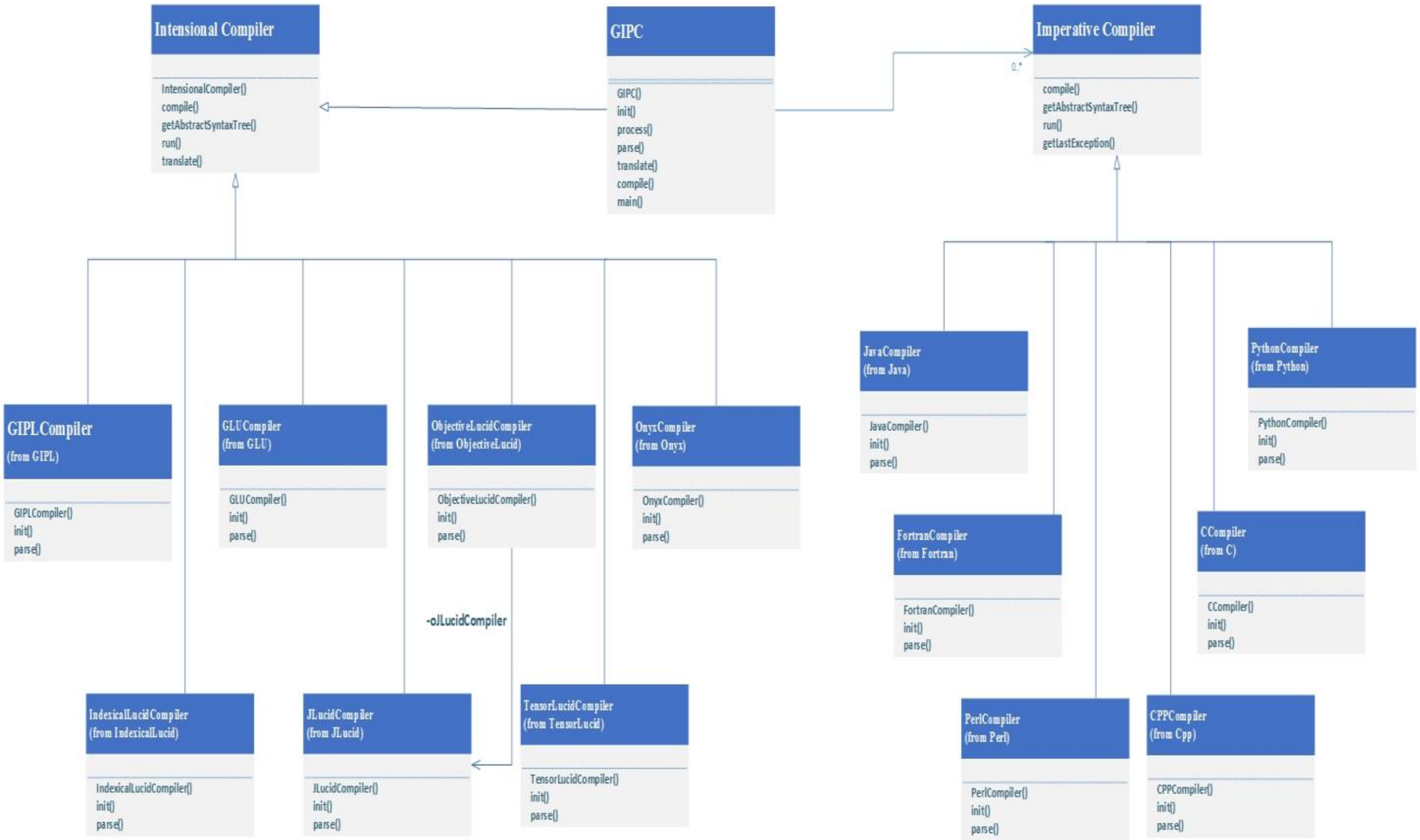

*Figure 12 GIPSY compiler (bulilt reading case studies)*

### *d) GIPC Class Diagram*

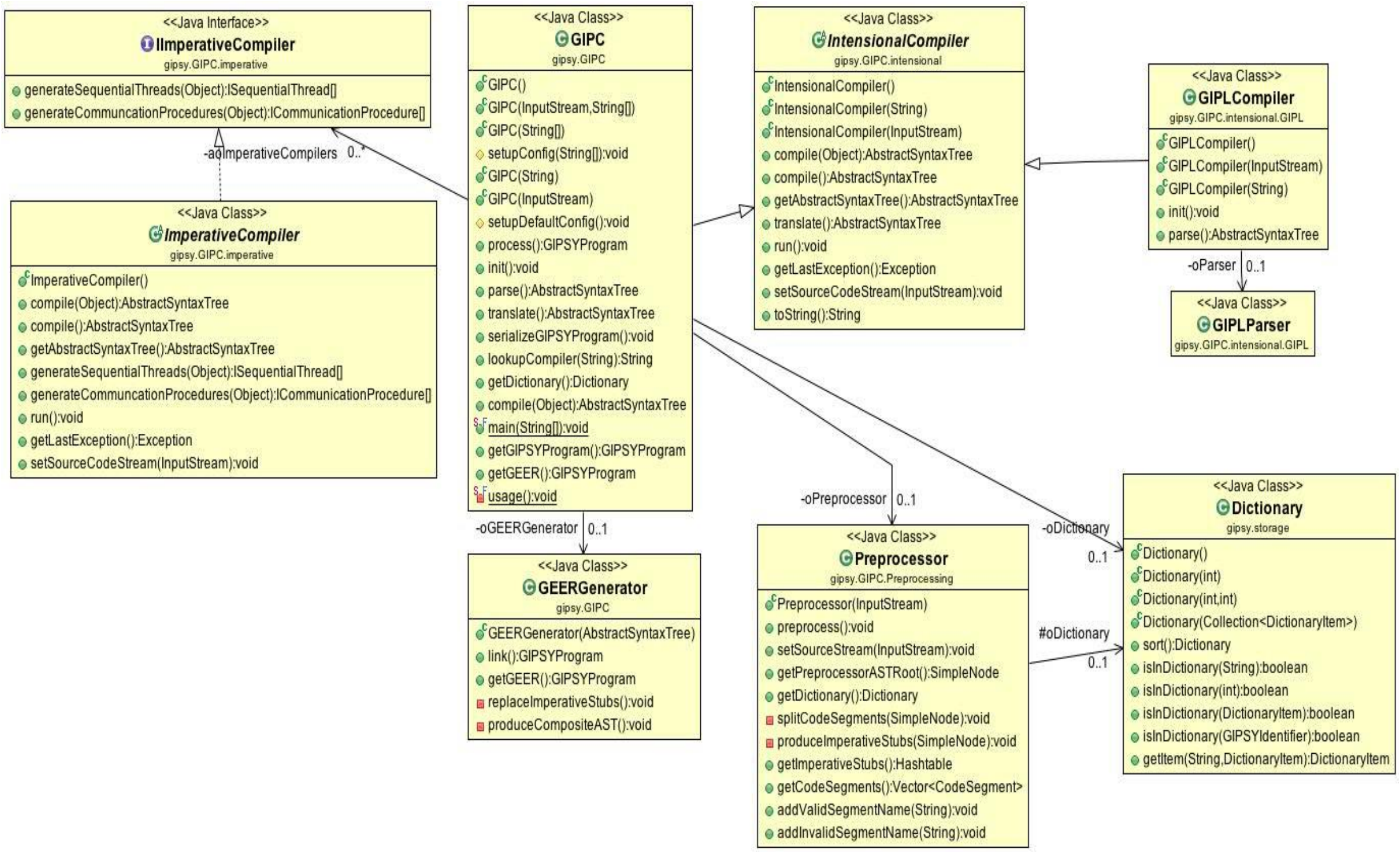

*Figure 13 GIPC Class Diagram*

**Description**

GIPSY compiler is composed of several compilers. Each compiler has its own language to compile. For instance, JLucid is compiled by JLucid Compiler and Forensic Lucid is compiled by Forensic Lucid Compiler. GIPC is divided into two types of compilers, Intensional Compiler and Imperative Compiler. All the Lucid compilers fall under the category of Intensional compilers. In GIPC, any Intensional Program is translated into the GIPSY understandable format i.e., GIPL (GIPSY Intensional Programming Language) using the **GIPLCompiler** and **GIPLParser**. GIPC compiles the code in a two stage process. The GIPSY intensional part is translated into C and this C program is compiled in the standard compilation method. Of the two parts in the source code the Lucid part defines data dependencies between variables and the sequential part that defines the granular sequential computation units. This Lucid part is compiled into a dependency structure due to the presence of dependencies. This dependency structure is named as Intensional Dependency Structure (IDS). The sequential functions defined in the second part of the GIPSY program are translated into imperative code using the second stage imperative compiler syntax, yielding imperative sequential threads. [32] The GIPC finalizes GIPSY program compilation by doing semantic analysis and eliminating Lucid functions and producing the demand AST (Demand Abstract Syntax Tree) along with linking in the generated STs (Syntax Trees) and CPs (Communication Procedures) from the imperative side. The GEER generator then produces the final linked version of a GIPSY program as a resource usable by the GEE (GEER). GIPC invokes Preprocessor on the incoming stream of GIPSY. PreProcessor does preliminary program analysis, processing, and splitting into chunks. Since a GIPSY program is a hybrid program consisting of different languages in one source file, there ought to be an interface between all these chunks. Thus Preprocessor constructs a preliminary Dictionary of symbols used in the whole program, after initial parsing and producing the initial parse tree. The AST and the Dictionary contain the generated accessor identifiers that are processed by the JLucid mechanisms.

### *e) GEE Class Diagram*

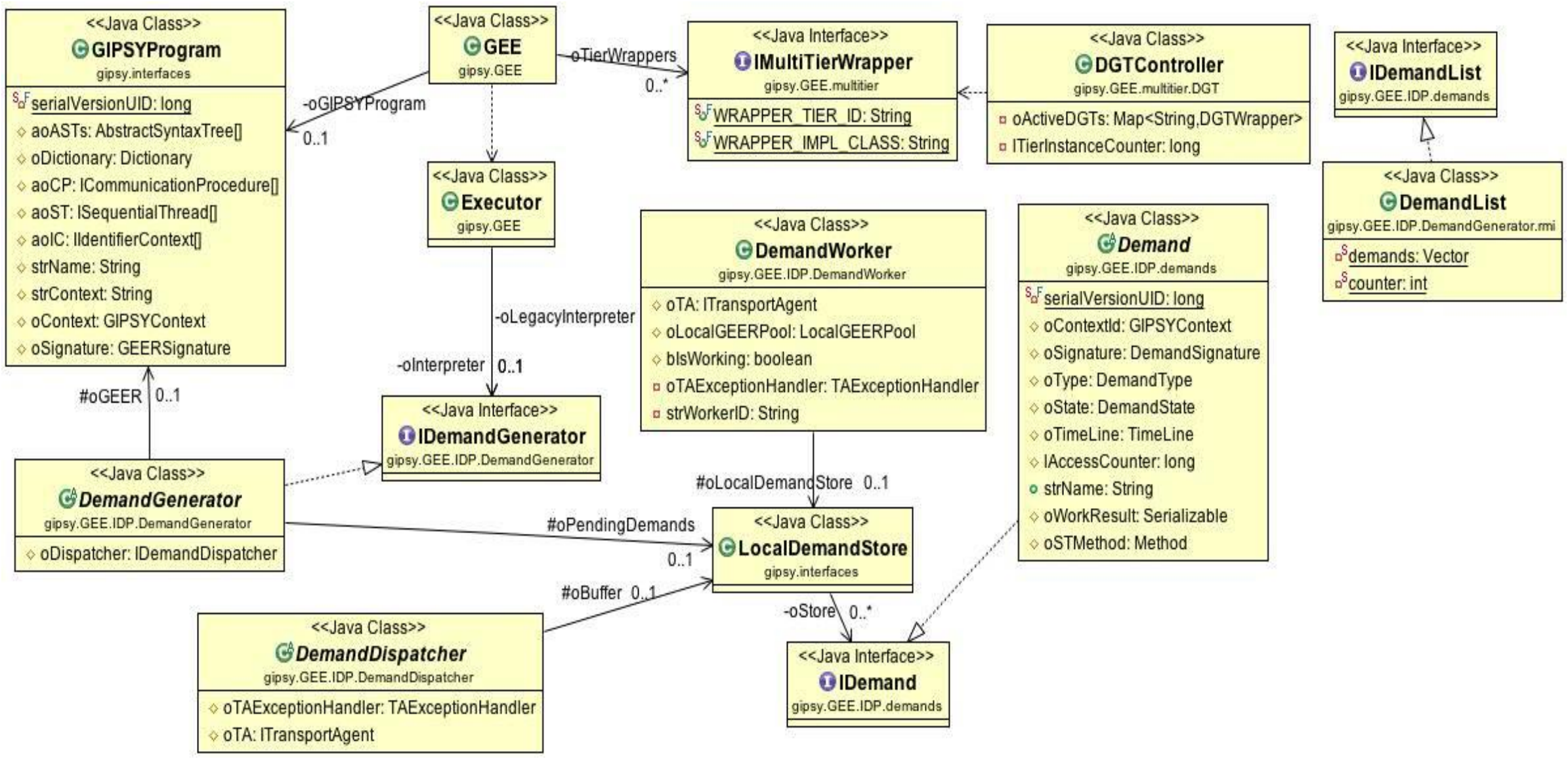

*Figure 14 GEE Class Diagram*

**Description**

**GEE** is the Eduction Engine for the demand driven architecture of GIPSY. GEE can be invoked either by GIPC GEER (represented as GIPSYProgram) or by the user providing **GIPSYProgram**. GEE loads this program and starts execution with the help of **Executor** thread. Executor generates the demands for the identifiers listed in the program and performs the calculation based on the results received while executing the program. **IDemandList** interface is consistently used by the **DemandGenerator** along with the **DemandDispatcherAgent** to be in agreement to the rest of the GEE. **LocalDemandStore** stores all the demands. DemandGenerator generates all the demands and it is part of the Demand Generator Tier. DemandGenerator has IVW (Intensional Value Warehouse) which stores all the demands. Additionally, DG also consist IDP (Intensional Demand Propagator). This IDP propagates the demands and helpful for the generation of demands. Few tiers are wrapped into **IMultiTierWrapper** interface. **DGTController** controls the demand stream. **DemandWorker** remains as the second major part of the GEE. DW consists of RIPE. RIPE is the runtime environment of GIPSY. It computes the RIPE sequential threads as demanded by the Generator or **DemandGenerator**.

*a) **Architecture Class Diagram***

**Description**

GIPSY has a demand driven and multitier architecture. GIPSY consists of four tiers. **NodeController** is an abstract class for each tier controller. Subclasses in it decide which tier type to instantiate. **NodeController** helps in adding new Tiers or TierController as GIPSY is a multitier. If necessary, there will be one controller for each tier that implements GenericTierWrapper running on every node. **NodeController** is responsible for adding or removing any tier instances from the node through **TierFatory** [22]. **TierFactory** is an instance of the abstract factory pattern. All the tier controllers have a reference to the subclass of TierFactory. This TierFactory creates objects of its respective tier type. All the GIPSY tiers were implemented in gipsy.GEE.multitier package. And this package consists of sub packages. Each sub package is of each tier type. The sub packages are gipsy.GEE.multitier.DGT.DGTWrapper, gipsy.GEE.multitier.DST.DSTWrapper, gipsy.GEE.multitier.DWT.DWTWrapper, gipsy.GEE.multitier.GMT.GMTWrapper. All these packages consists of Wrapper classes. The Wrapper classes are **DGTWrapper, DSTWrapper, DWTWRapper, GMTWrapper**. The Wrapper classes are inherited under one class **GenericTierWrapper**. This WrapperClass implements the most common functionality of **IMultiTierWrapper** interface which also resides in the gipsy.GEE.multitier package. Both **DGTWrapper** and **DWTWrapper** classes have the common oGEERPool data member, which is a collection of objects of type **GIPSYProgram** (instances of GEER). The GEERPool acts like a local cache containing GEERs that the tier can compute demands for [22]. GIPSYProgram is the container of AST (Abstract Syntax Tree) and the Dictionary of identifiers. GEERPool consists of GEERs as a cache in each of DGT and DWT. DSTWrapper is similar to DGTWrapper and DWTWrapper, except when it inherits from the base class GenericTierWrapper, the DSTWrapper encapsulates oStorageSubsystem, which is an object of type IVWInterface. **IVWInterface** is an integrated Intensional Value Warehouse also referred as DS (Demand Store).

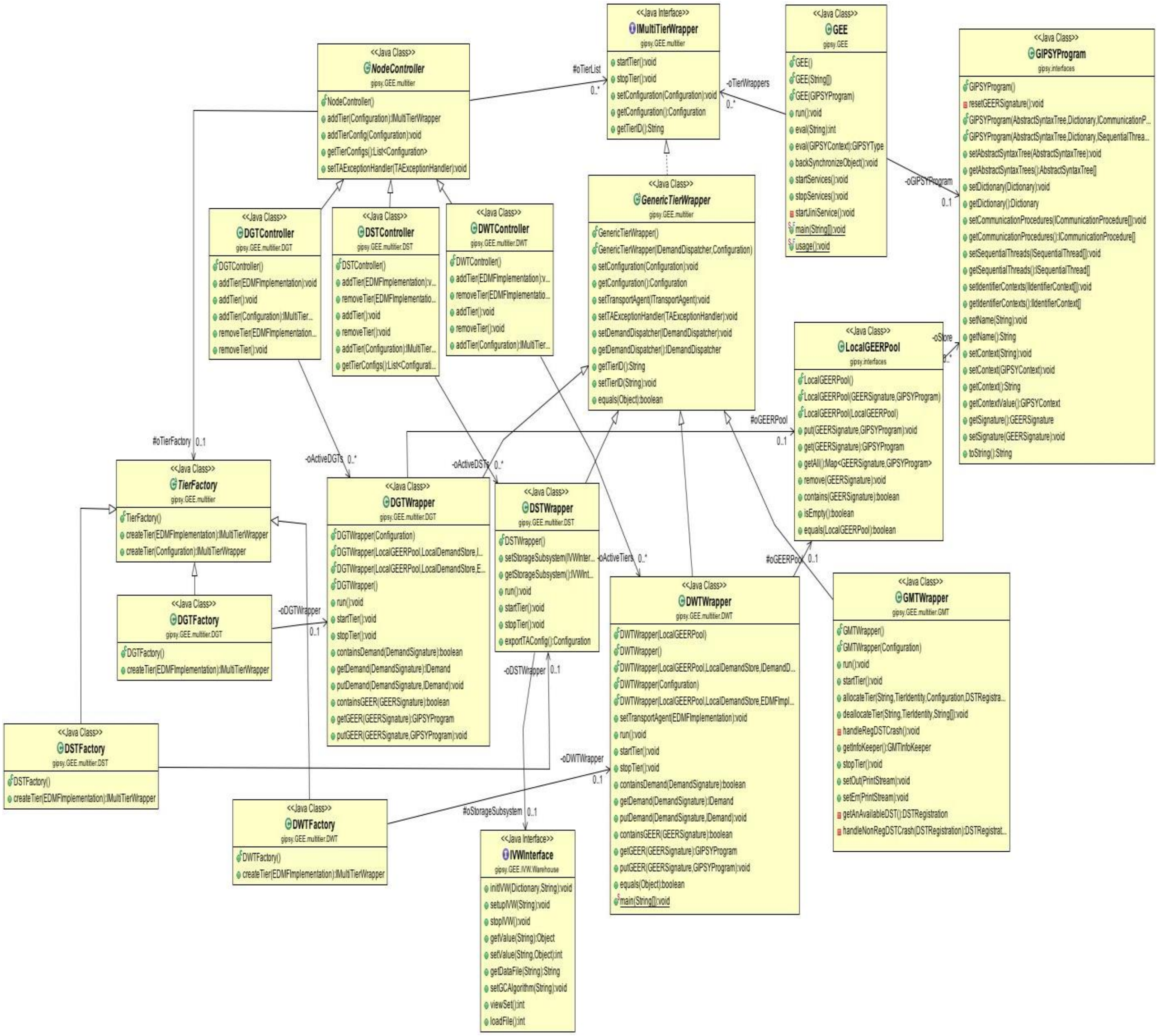

*Figure 15 Architecture Class Diagram*

**ObjectAid**

objectAid is UML Explorer is an agile and lightweight code visualization tool for the Eclipse IDE. It can be installed in eclipse IDE by adding objectAid plugin. It is able to show the source code and the libraries in the classes used. UML class, Sequence diagrams can be built automatically and also the relations can be associated inside automatically.

**Differences between the concepts and the actual classes – GIPSY**

| Domain Model | Class Diagram |
|---|---|
| IntensionalProgram | Lucid |
| GIPSYCompiler | GIPC |
| EductionEngine | GEE |
| RuntimeEnvironment | RIPE |
| Resource | GEER |

The differences found are mentioned in the above table.

i) IntensionalProgram is named the concept of input provided to the GIPSY. The intensional Program in major consists of Lucid Programs.

ii) GIPSYCompiler is considered the real concept for the GIPSY compiler i.e., GIPC (General Intensional Programming Compiler).

iii) EductionEngine is the General Eduction Engine, one of the major module of GIPSY

iv) RuntimeEnvironment is the RIPE (Runtime Intesional Programming Environment). It is used to compute Sequential threads demanded by the Demand Generator

v) Resource is the .gipsy file produced by GIPSY compiler. It is called as the GEER (GEE Resource) in the GIPSY.

**Mapping conceptual classes and actual systems - GIPSY**

- GIPSYCompiler to GIPC

The GIPSYCompiler is the compiler of GIPSY. Its name is GIPC (General Intensional Programming Compiler). In general, this compiler is the abstract of all the compilers in GIPSY.

- IntesionalProgram to LucidProgram

The Intensional Program is the input provided to the user. As the Lucid Program is intensional in nature, Lucid program is considered here as the Intensional Program.

- EductionEngine to GEE

EductionEngine is the GEE (General Eduction Engine) which produces demands accepting resources from the GIPC

- RuntimeEnvironment to RIPE

RuntimeEnvironment is the RIPE (Runtime Intensional Programming Environment). It is invoked by GIPC

- Resource to GEER

Resource is the file produced by GIPC and called as the GEER (GEE Resource).

**Two Classes and the relationship between them - GIPSY**

There are classes having the same attributes, methods. GEE class and RIPE are related in GIPSY.
Common method is *usage()*

```
Public class GEE extends StorageManager implements
Runnable
{
if(this.oOptionProcessor.isActiveOption(OPT_HELP))
                {
                        usage();
                }
public static final void usage()
        {
        }
```

**RIPE Class**

```
Public class RIPE extends BaseThread
{
public static final void usage()
        {
        }
```

**Two Classes and the relationship between them – DMARF**

**FeatureExtraction:**

```
public abstract class FeatureExtraction
extends StorageManager
implements IFeatureExtraction
{
   public boolean extractFeatures()
           throws FeatureExtractionException
           {
       ...
     }
}
```

**Preprocessing:**

```
public abstract class Preprocessing
extends StorageManager
implements IPreprocessing
{
   public boolean preprocess()
           throws PreprocessingException
           {
     ......
     }
}
```

## V. REFACTORING

### A. Identification of Code Smells & System Level Refactorings

#### *1) DMARF*

**Code Smell # 1**

In MARF.java class, Long class and GOD Object exist.

*Solution:*

- We moved recognize() and recognize(Sample) methods to a new class named “MARFProduct.java” for high-cohesion and low-coupling.
- We created the instance of the newly created class in MARF.java to avoid GOD object.
- Then we called the recognize() method from the instance of the newly created class in MARF.recognize().
- We changed the access levels of the required methods and fields.

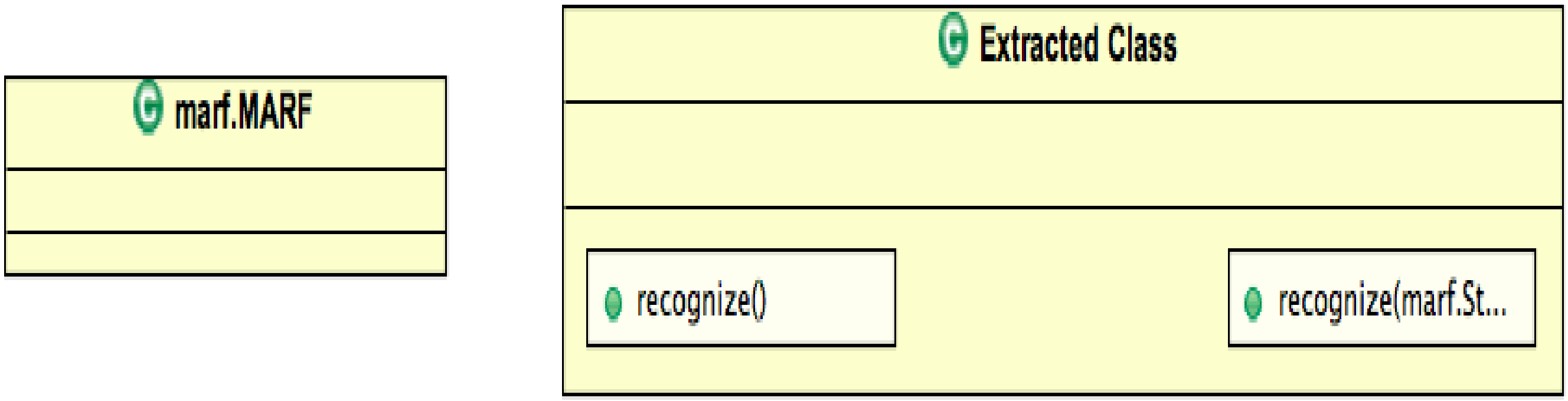

*Figure 16UML for MARF.java and MARFProduct.java after solution*

**Code Smell # 2**

In Parser.java class, Feature envy exists as skipErrors() in parser.java is using features of LexicalAnalyzer.java.

*Solution:*

- We moved the skipErrors() method to LexicalAnalyzer.java to avoid feature envy.
- We changed the invocation of moved() method by calling it through the instance of LexicalAnalyzer.java.

2) ***GIPSY***

**Code Smell # 1**

In GIPC.java class, protected methods are used that result in low-cohesion and high-coupling. Plus long sequences of method calls are also present.

*Solution:*

- First, we created a new class named GIPCSetupConfig.java.
- In this class we moved two methods, setupConfig and setupDefaultConfig from GIPC.java.
- We changed the types of these methods from protected to public for high-cohesion and low-coupling.
- We created a field in GIPC.java holding a reference of GIPCSetupConfig.java namely gIPCSetupConfig.
- We changed the access of extracted member from “setupDefaultConfig()” to “gIPCSetupConfig.setupDefaultConfig()”.
- Similarly, we are accessing all extracted members using object of the newly created class.

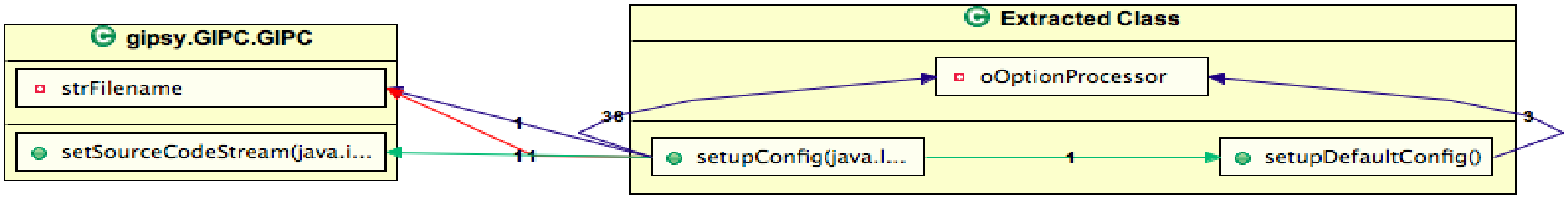

*Figure 17UML for GIPC.java and GIPCSetupConfig.java after solution*

**Code Smell # 2**

In LegacyInterpreter.java class, Feature Envy is caused due to inappropriate method invocation.

*Solution:*

- First we moved the contents of eval() method into SimpleNode.java class by creating a new method “eval()”.

- In LegacyInterpreter.execute() method, we have changed the invocation of moved method to reduce feature envy.
- In SimpleNode.java, we added the required import declarations.

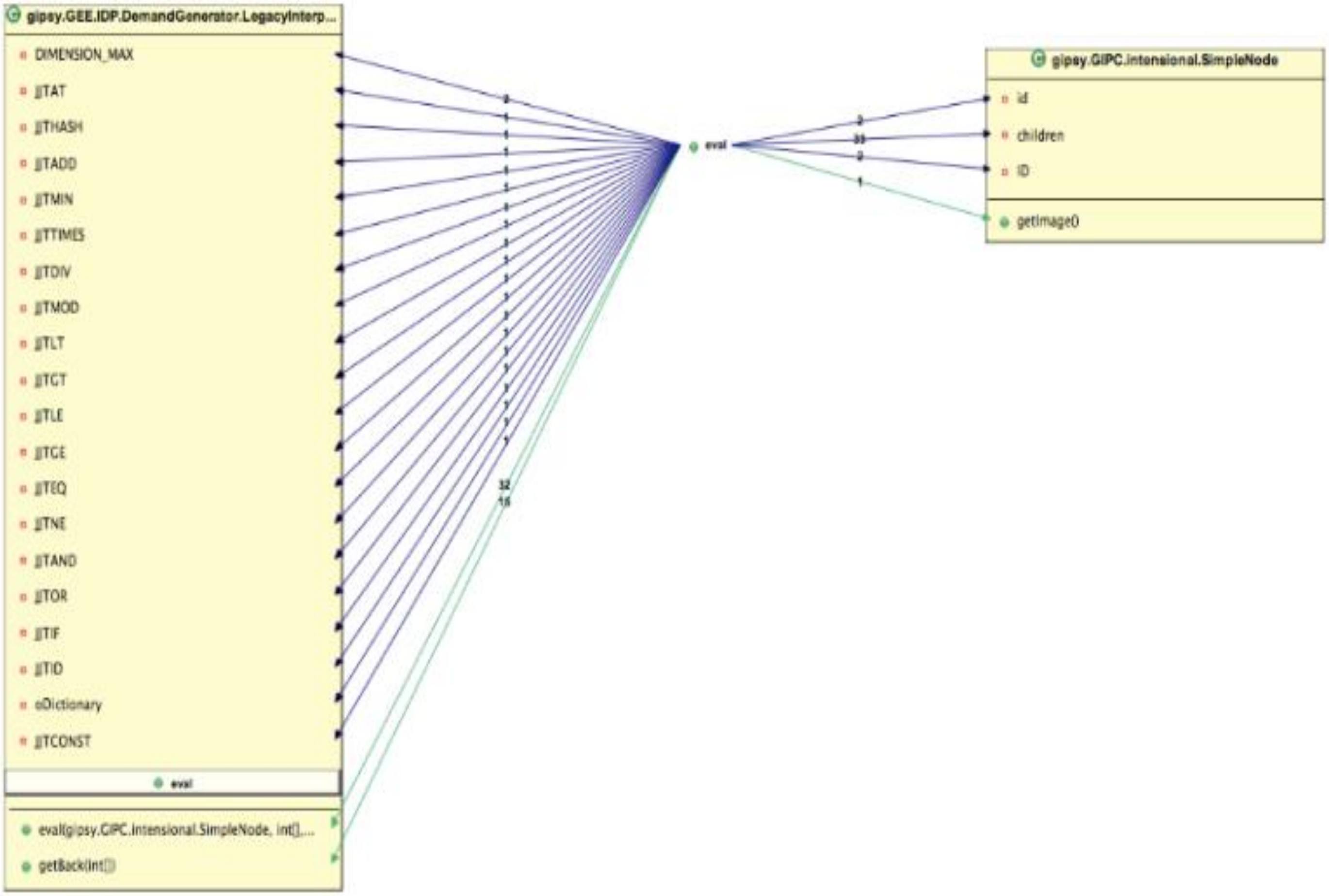

*Figure 18 UML for LegacyInterpreter.java and SimpleNode.java after solution*

## B. Specific Refactorings to Be Implemented

### 1) *DMARF*

**Refactoring # 1**

A method named create() in FeatureExtractionFactory.java contains a complex switch case. In order to refactor this switch case we came up with the following approach:

1. We can create an interface which will have a build() method.
2. For each value the build() method will return a different object.
3. We will create a HashMap to map the relevant resultant object according to the provided input value.
4. Switch statement will not be required in create() method as the relevant object will be fetched from HashMap according to the input.

```
public final class FeatureExtractionFactory{

    public static final IFeatureExtraction create(final int piFeatureExtractionMethod, IPreprocessing poPreprocessing) throws FeatureExtractionException{}
}
```

**Test Case:**
In order to verify the correct functionality of this method we will create a class named FeatureExtractionFactoryTest.java in which create() method will be tested using some dummy data.

**Refactoring # 2**

A method named serialize() in LexicalError.java, SemanticError.java and SyntaxError.java is similar in all the three classes. In order to refactor this method we came up with the following approach:

1. We moved the serialize() method into their existing parent class CompilerError.java to avoid duplication.
2. The type of the error, that is being checked inside the method, will be done by checking the type of the instance.
3. We will also build the string dynamically to identify the type of errors separately. For Example: Syntax Error or Semantic Error.

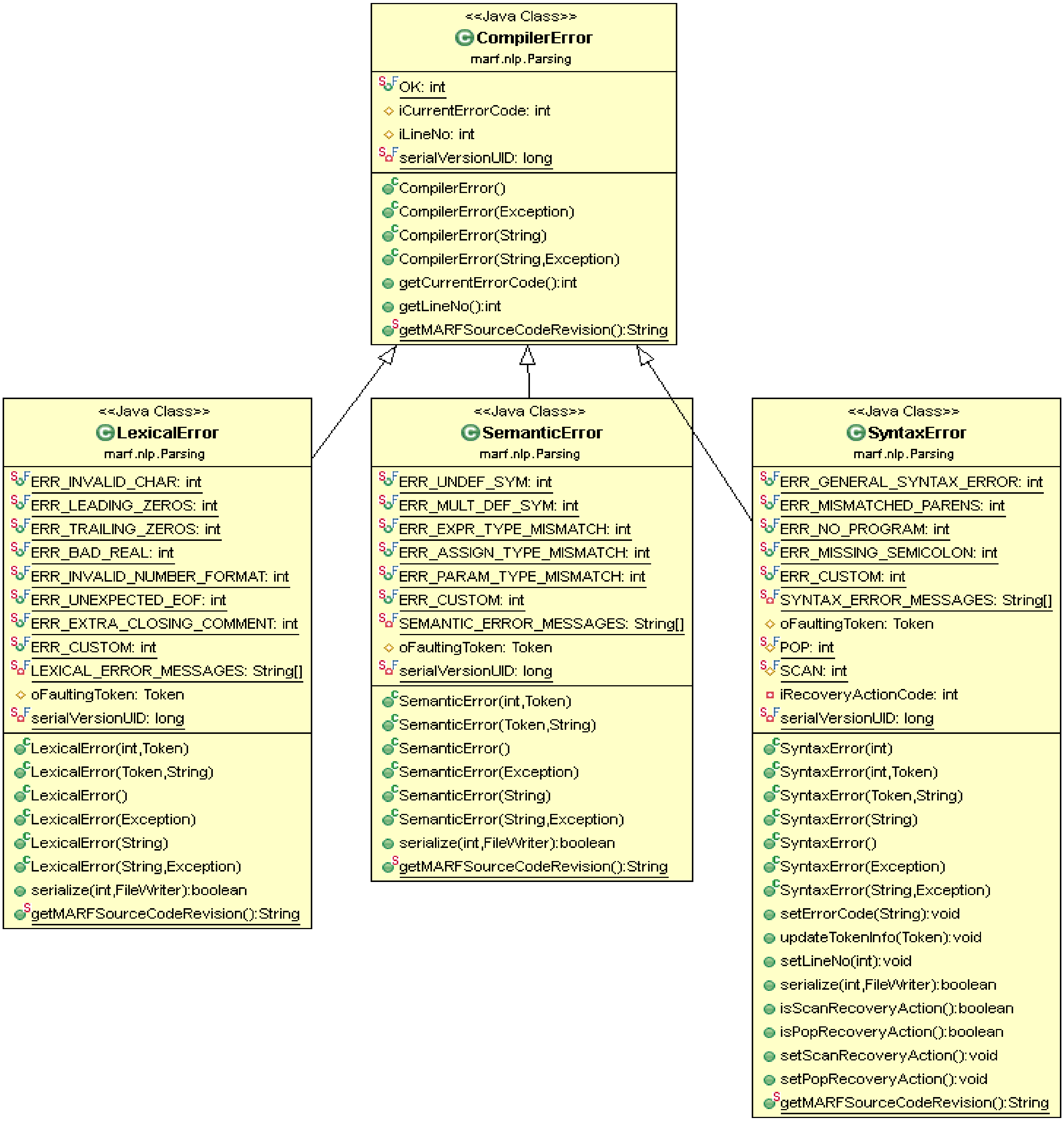

*Figure 19 UML for relevant classes before moving method*

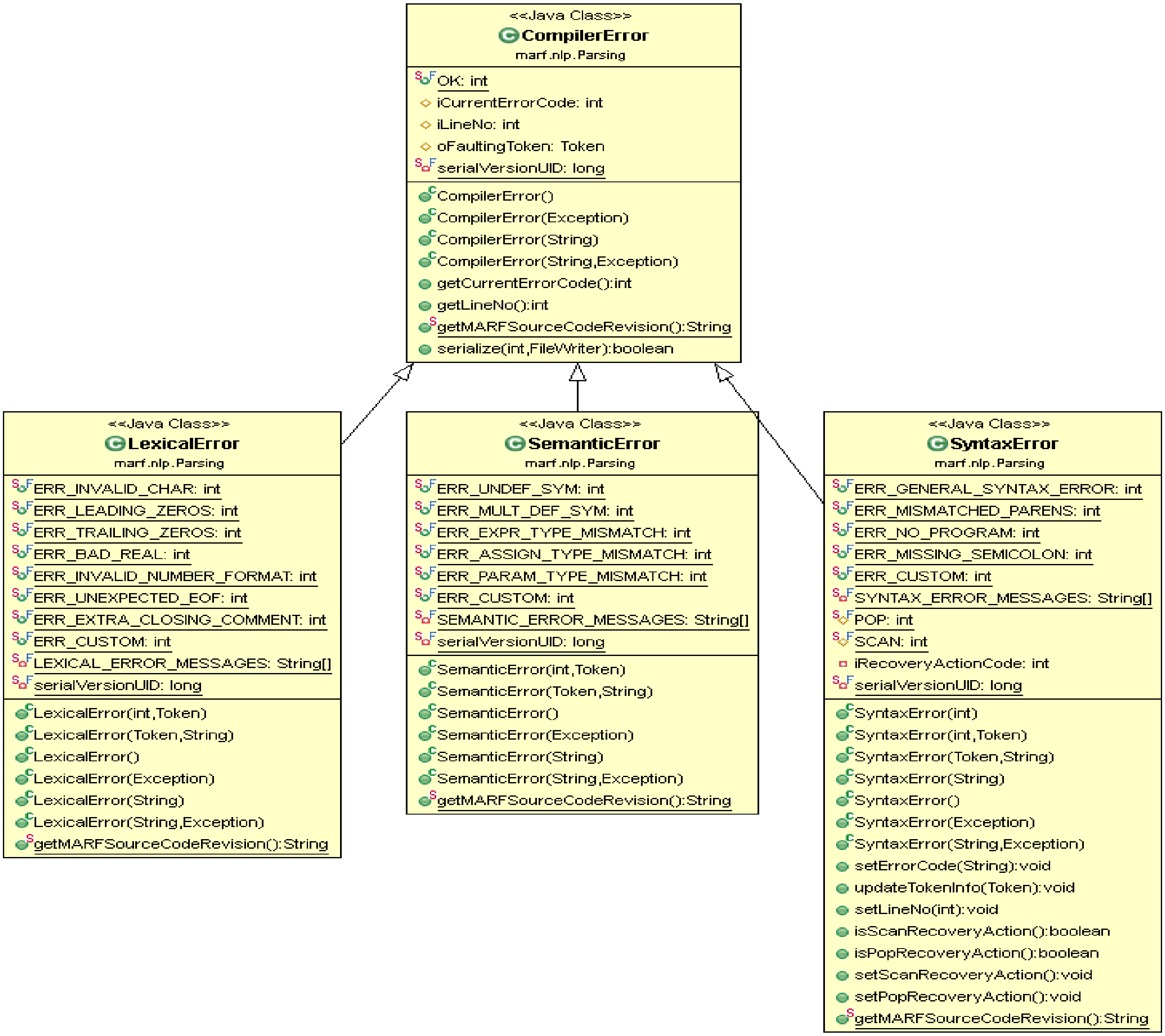

*Figure 20 UML of relevant classes after moving method*

**Test Case:**

Test cases will be implemented for the above mentioned classes named CompileErrorTest.java using some dummy data in order to verify correct execution of serialize() method invoked via a subclass object.

2) *GIPSY*

**Refactoring # 1**

- Method named execute(Dictionary poDictionary, GIPSYContext poDimensionTags) has been used in both Interpreter.java and LegacyInterpreter.java.
- Similarly, the method eval(SimpleNode poRoot, GIPSYContext[] paoContext, int piIndent) and getBack(GIPSYContext[] paoOriginalContext) has also been used in both Interpreter.java and LegacyInterpreter.java.
- In order to refactor this, we removed execute(Dictionary poDictionary, GIPSYContext poDimensionTags), eval(SimpleNode poRoot, GIPSYContext[] paoContext, int piIndent) and getBack(GIPSYContext[] paoOriginalContext) method from LegacyInterpreter.java to reduce redundancy.

- We extended LegacyInterpreter.java by Interpreter.java.

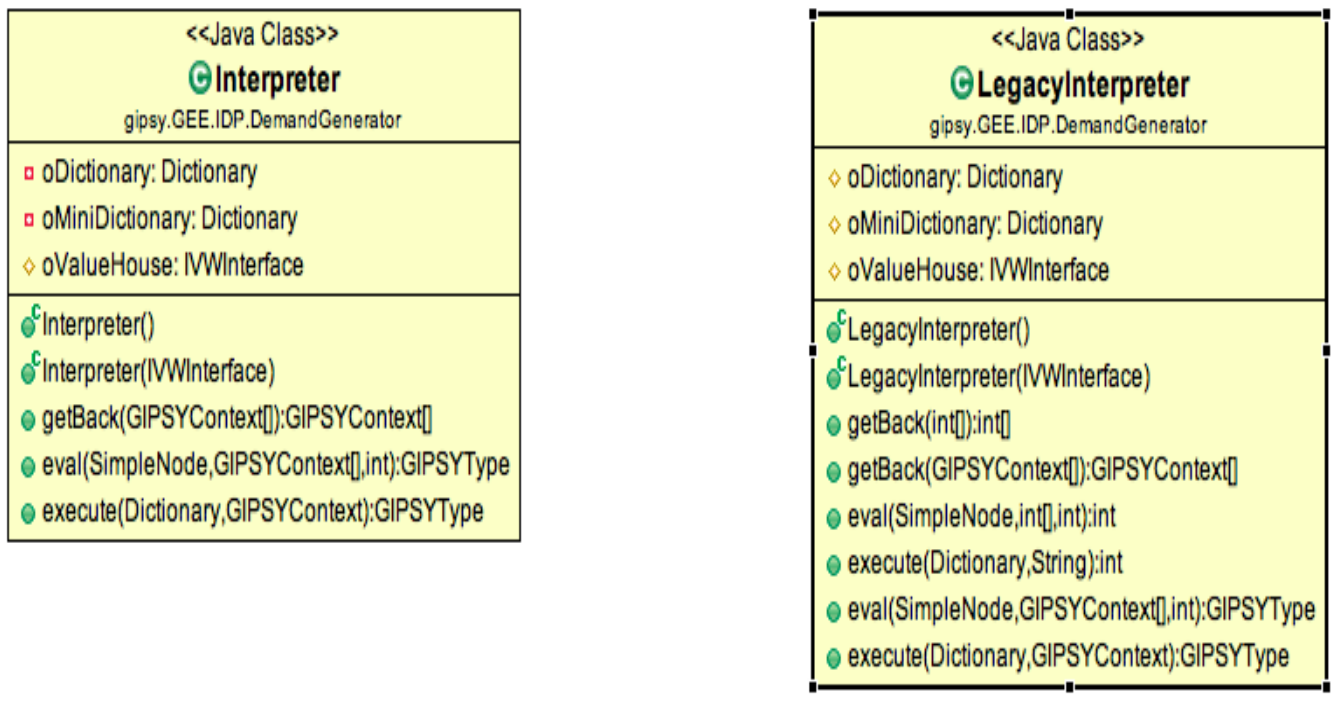

*Figure 21 UML of relevant classes with method duplication*

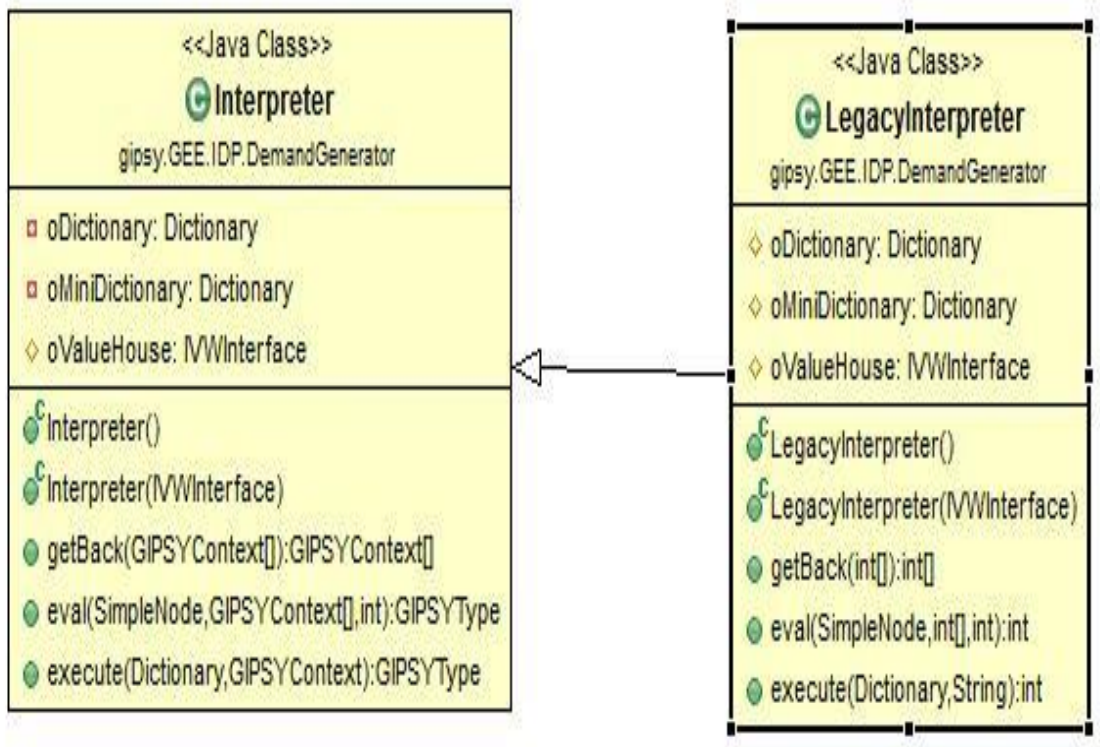

*Figure 22UML of relevant classes without method duplication*

## VI. IDENTIFICATION OF DESIGN PATTERNS

### A. DMARF

#### 1) *Composite Pattern*

A composite pattern is the recursive structure in an inheritance tree, where one of the leaves is an aggregation of several roots. This pattern can be discovered in our graph. [25]

The reason behind the use of this pattern is that the group of objects should be used as the single object so that the client can access the hierarchy through the components and they are not aware if they are dealing with composite or the leaf node.

In our case we have one interface (Composite), ASSLEVENTCACHER.java, which contains a method that is used by three classes (Leaf nodes) that are ASSLFLUENT.java, ASSLRECOVERY_PROTOCOL.java, ASSLEVENT.java by implementing that interface.

**Class and Method Declarations:**

```
public interface ASSLEVENTCATCHER
{
        public void notifyForEvent ( ASSLEVENT poEvent );
}

public class ASSLFLUENT extends Thread
        implements  ASSLEVENTCATCHER
{
        public synchronized void notifyForEvent (
        ASSLEVENT poEvent )
        {
                …
        }
}
public class ASSLRECOVERY_PROTOCOL
        implements  ASSLEVENTCATCHER
{
        public synchronized void notifyForEvent (
        ASSLEVENT poEvent )
        {
                …
        }
}
public class ASSLEVENT
        extends Thread
        implements ASSLEVENTCATCHER,
        ASSLMESSAGECATCHER
{
        public synchronized void notifyForEvent (
        ASSLEVENT poEvent )
        {
                …
        }
}
```

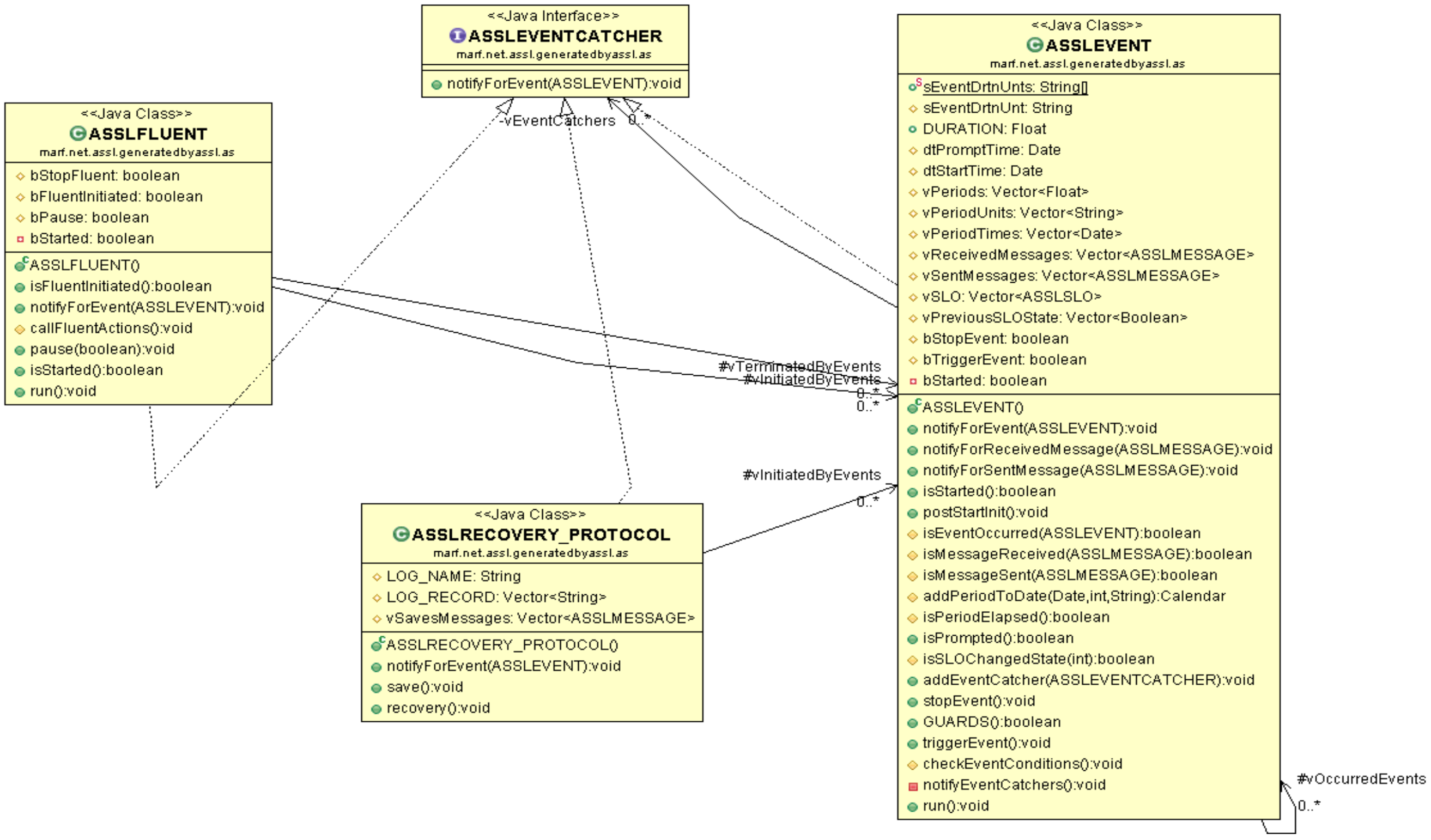

*Figure 23 UML of interacting classes having Composite Pattern*

### 2) *Adapter Pattern*

The adapter design pattern is used to convert the interface of a class into another interface that is expected by its clients. The adapter design pattern allows classes to cooperate that otherwise would be incompatible due to the differences in expected interfaces [28].

As DMARF is currently implemented for RMI and CORBA so there is a need of an adapter, which is responsible for translating MARF data structures to CORBA, For example there are some methods available in MARFObjectAdapter.java that loads configuration of MARF with compatibility for RMI and CORBA.

**Class and Method Declarations:**

```
public class MARFObjectAdapter
{
	public static final marf.net.server.corba.Storage.Configuration getCORBAConfiguration (marf.Configuration poMARFConfig)
	{
	…
	}
}
public class SpeakerIdentCORBAClient
implements ICORBAClient
{
	public /*static*/ final void start(String[] argv)
	{
	…
	}
}
```

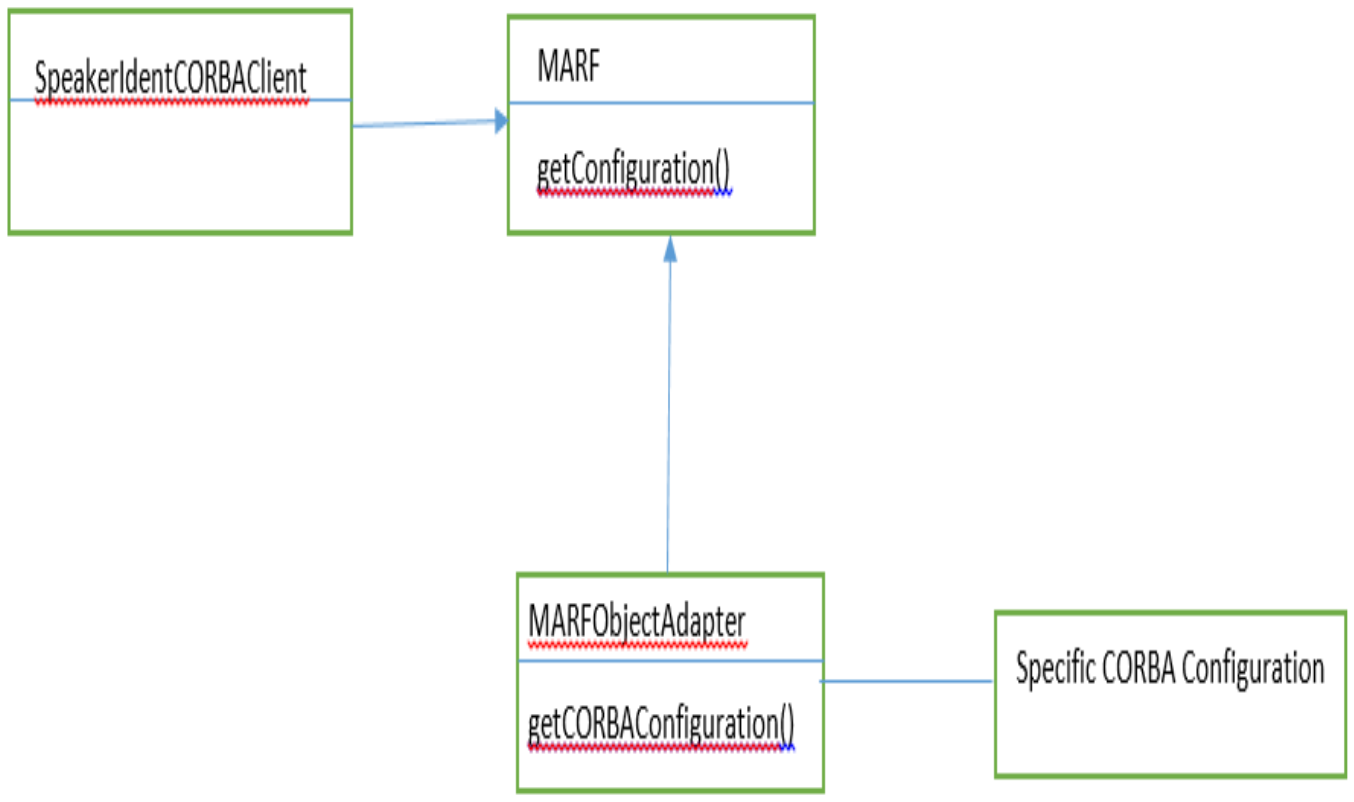

*Figure 24 UML of interacting classes having AdapterPattern*

### 3) *Factory Pattern*

The dynamic factory pattern describes a factory that can create product instances based on concrete type definitions stored as external metadata. This facilitates adding new products to a system without having to modify code in the factory class [30].

In DMARF this pattern has been used in FeatureExtractionFactory.java class as there are different types of feature extraction processes having separate class for each type, so the purpose of the factory pattern is to create correct type of instance which is required during pipeline stage.

```
public final class FeatureExtractionFactory
{
        public static final IFeatureExtraction create(final int piFeatureExtractionMethod, IPreprocessing poPreprocessing)
        throws FeatureExtractionException
        {
        ……
        }
}
```

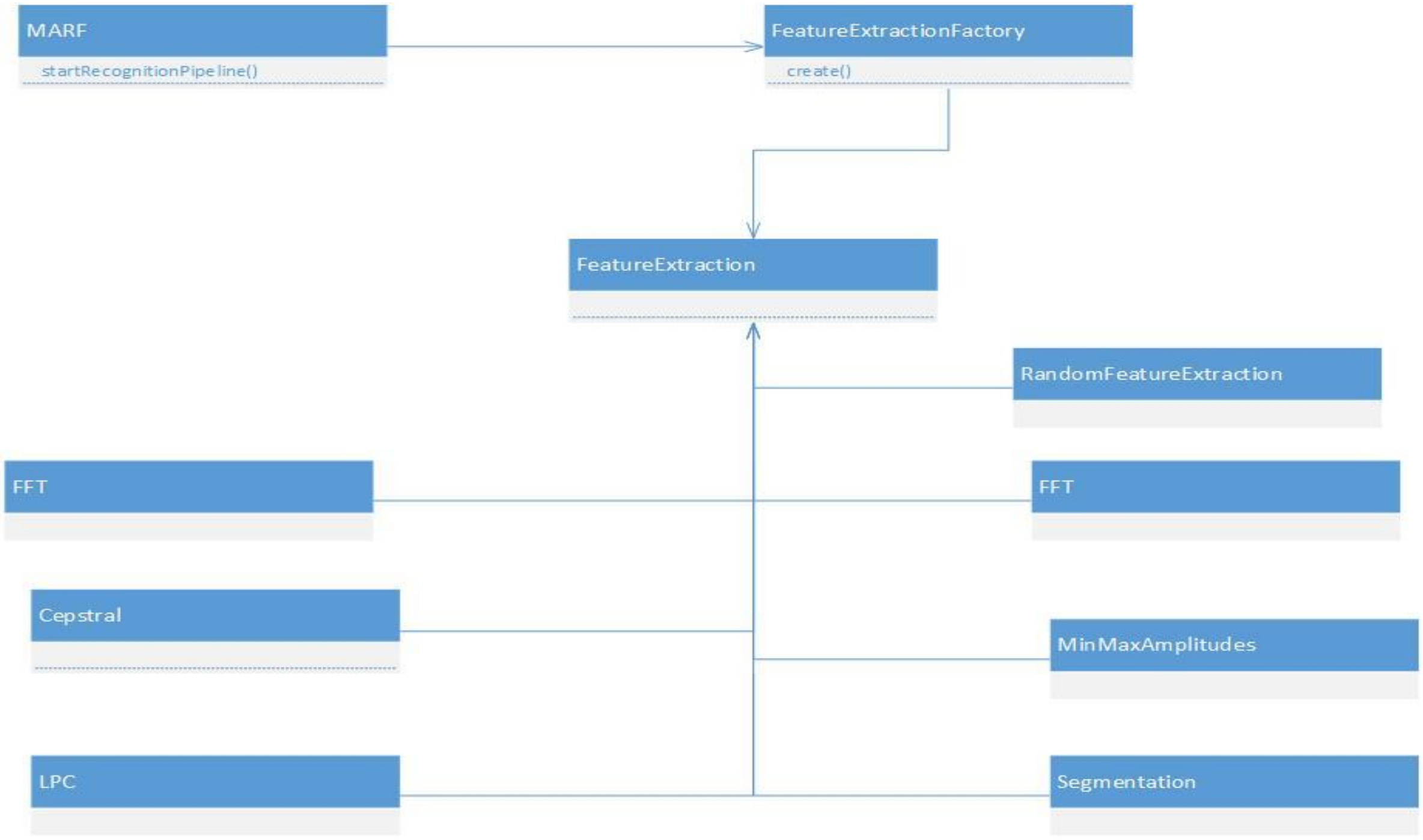

*Figure 25 Factory Pattern*

## B. GIPSY

### 1) *Singleton Pattern*

This is a creational pattern that "ensures a class only has one instance", with a single global access point. (A good example of a situation where this might be appropriate is for a spooler object managing a printer [29].

In GIPSY there is a class named DemandPool.java where this pattern has been used in order to have only one instance of it so that the worker will fulfill this demand.

```
public class DemandPool
{
        public static synchronized DemandPool getInstance()
        {
        ….
        }
}
```

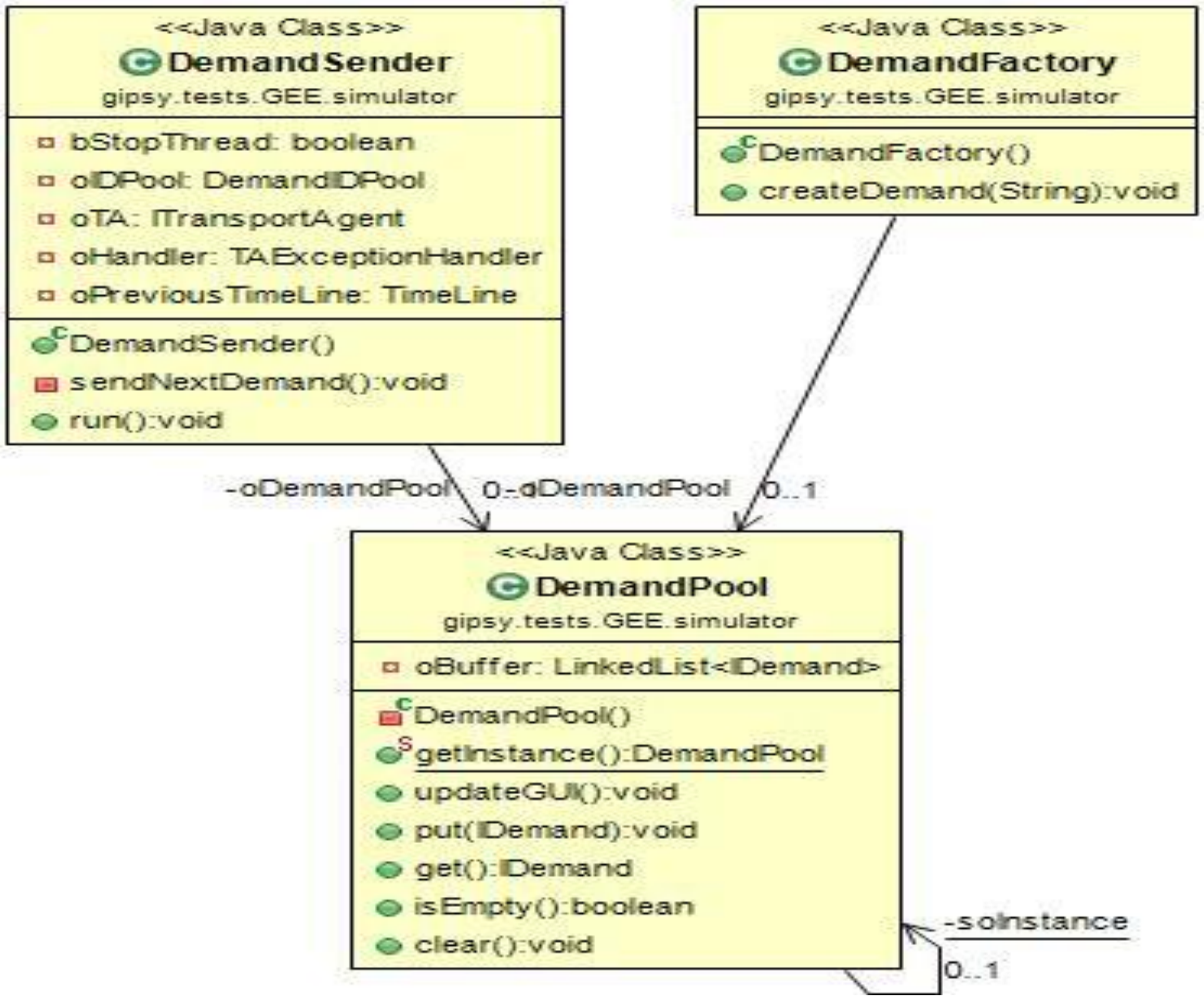

*Figure 26 Singleton Pattern*

### 2) *Facade Pattern*

The Facade pattern is used to define a higher-level interface to a subsystem that makes the subsystem easier to use. This pattern is very useful where subsystems provide collections of related features that can be grouped together and presented in a unified way using a Facade class [31].

There is a class named GIPSYNode in gipsy project that is representing the facade pattern as it is providing a single point of contact to create a GIPSY Node instance based on configuration, load all configuration files and assign them to their corresponding controllers.

```
public class GIPSYNode
extends Thread
{
        public GIPSYNode(Configuration poNodeConfig)
        {
        aoControllers[0] = new GMTController();
        aoControllers[1] = new DSTController();
        aoControllers[2] = new DGTController();
        aoControllers[3] = new DWTController();
        ….
        }
}

public class GIPSYNodeTestDriver
{
        public static void main(String[] pastrArgs)
        {
        GIPSYNode oNode = new
        GIPSYNode(oNodeConfig);
        …..
        }
}

public class GIPSYGMTController
{
        public void startGMTNode() throws
        MultiTierException
        {
        GIPSYNode oNode = new
        GIPSYNode(oNodeConfig);
        }
}
```

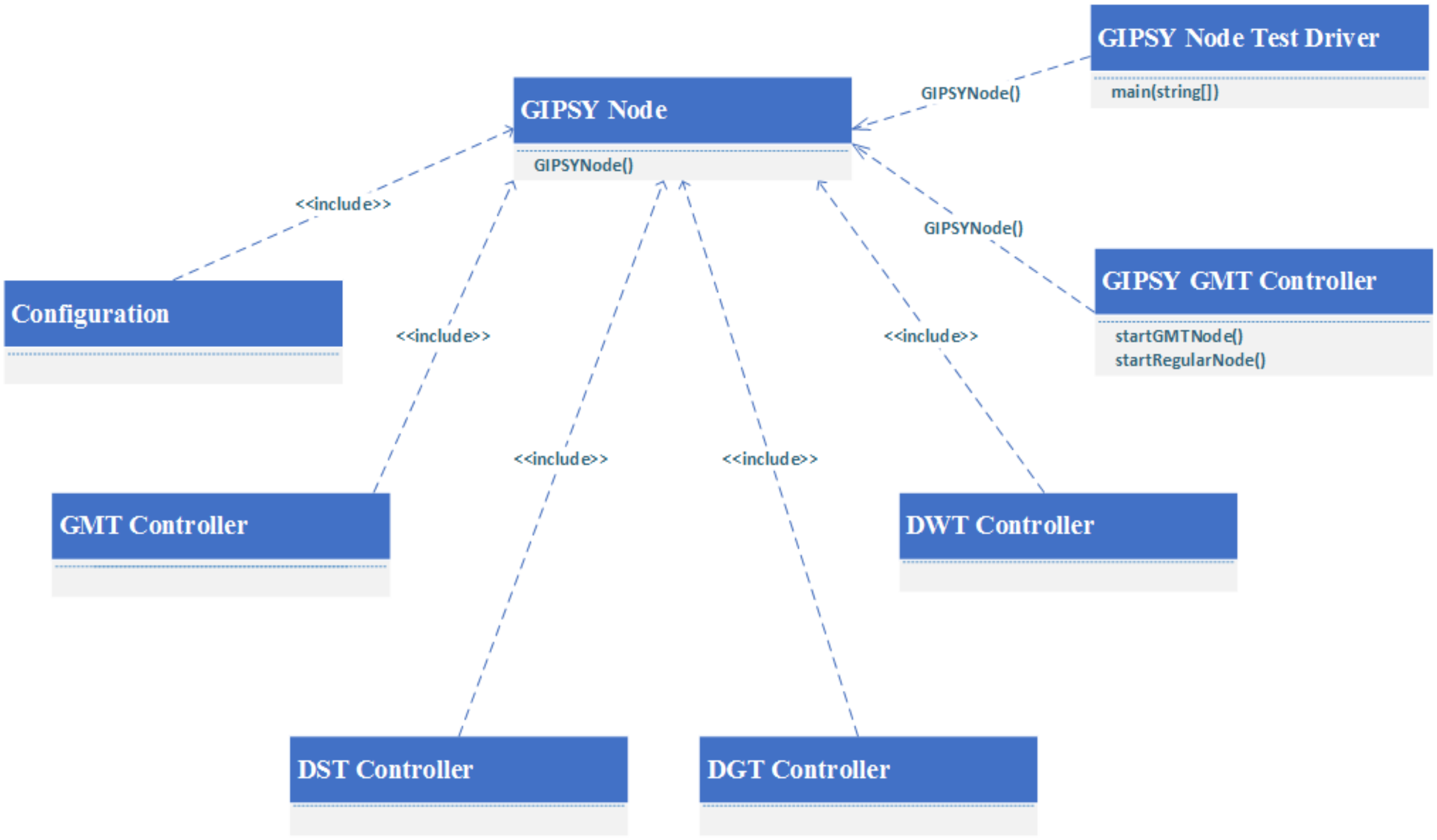

*Figure 27 Facade Pattern*

### 3) *Observer Pattern*

The observer pattern defines interactions between two types of objects, observers and subjects, where the observers must react to changes in the states of the subjects [32]. Here this pattern is being used to indicate other classes that are observing it to draw popup menu on screen according to selected indexes.

```
public static class GIPSYTiersMenu extends JPopupMenu
{
        ….
}

public interface GIPSYTierMenuListener<V>
{
  void setGIPSYTierAndView(V v, VisualizationViewer visView);
}

public static class CreateGIPSYTierPropertyItem extends JMenuItem implements
        GIPSYTierMenuListener<GIPSYTier>, MenuPointListener
{
        ….
}
public static class AllocateGIPSYTier<V> extends JMenuItem implements GIPSYTierMenuListener<V>
{
        ….
}

public static class DeAllocateGIPSYTier<V> extends JMenuItem implements
        GIPSYTierMenuListener<V>
{
        ….
}
```

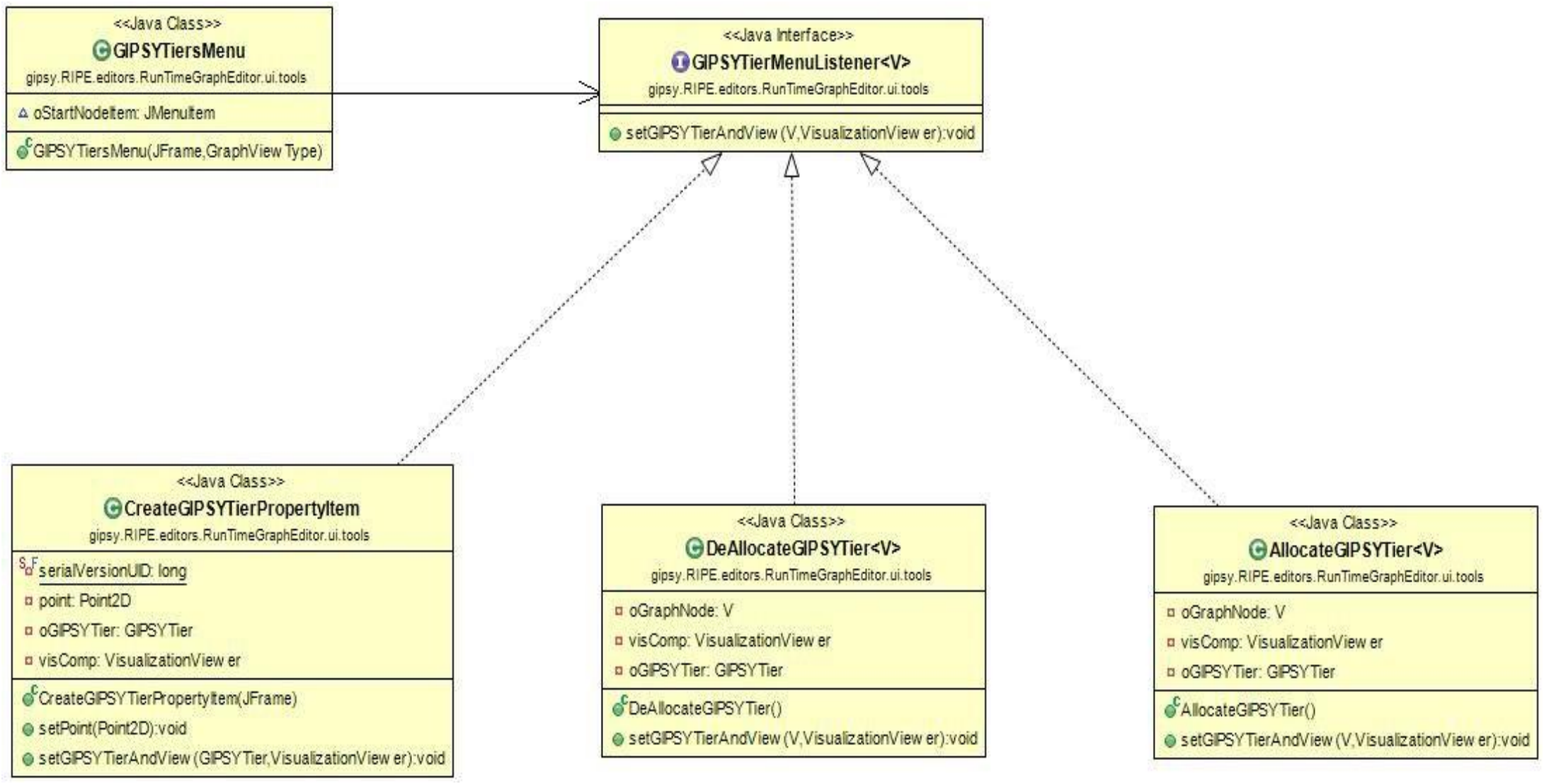

*Figure 28 Observer pattern*

### C. Tool Used

#### 1) *UML Lab Modelling IDE*

It is the first Eclipse based modelling IDE that combines modelling and coding with an intuitive UML diagram editor and next generation round trip engineering. Template-based reverse engineering allows you to use the same Xpand[27] Templates for forward and reverse engineering [26].

PATTERNS IDENTIFICATIONS BY TEAM MEMBERS:

| Design Pattern | Team Member |
|---|---|
| Factory Pattern | Shahroze Jamil |
| Singleton Pattern | Osama Yawar |
| Composite Pattern | Muhammad Nadir |
| Adapter Pattern | Tahir Ayoub |
| Facade Pattern | Yashwant Beeravelli |
| Observer Pattern | Parham Darbandi |

*Table 8 Individual Pattern Identification of GIPSY and DMARF*

## VII. IMPLEMENTATION

### A. Refactoring changesets and Diffs

#### 1) *DMARF*

##### a) *Refactoring # 1:*

**Change 0 / 2: Create a new Interface IFeatureExtractionProvider**

We created an interface named IFeatureExtractionProvider.java in which a method named build() has been declared that will be overridden in FeatureExtractionFactory.java in order to create and save different types of objects in a map.

```
### Eclipse Workspace Patch 1.0
#P marf
Index:
src/marf/FeatureExtraction/IFeatureExtrac
tionProvider.java
=========================================
==========================
RCS                                  file:
src/marf/FeatureExtraction/IFeatureExtrac
tionProvider.java
diff                                    -N
src/marf/FeatureExtraction/IFeatureExtrac
tionProvider.java
--- /dev/null     1  Jan  1970  00:00:00  -
0000
+++
src/marf/FeatureExtraction/IFeatureExtrac
tionProvider.java 1  Jan  1970  00:00:00  -
0000
@@ -0,0 +1,7 @@
+package marf.FeatureExtraction;
+
+import marf.Preprocessing.IPreprocessing;
+
+public interface
IFeatureExtractionProvider {
+     public IFeatureExtraction
build(IPreprocessing poPreprocessing);
+}
```

**Change 1 / 2: Remove complex switch case statement from FeatureExtractionFactory class**

Previously, a complex switch case statement was used in create() method in order to create the relevant object based on value of piFeatureExtractionMethod parameter which is being passed into it, now we have replaced this with HashMap by storing all the relevant types of objects in it.

```
/**
 * Instantiates a FeatureExtraction
module indicated by
```

```
* the first parameter with the 2nd
parameter as an argument.
@@ -95,73 +181,17 @@
* @see MinMaxAmplitudes
* @see FeatureExtractionAggregator
*/
+
public static final IFeatureExtraction
create(final int
piFeatureExtractionMethod, IPreprocessing
poPreprocessing)
      throws FeatureExtractionException
      {
-     IFeatureExtraction
      oFeatureExtraction = null;
-
-
      switch(piFeatureExtractionMethod)
-     {
-     case MARF.LPC:
-
      oFeatureExtraction = new
LPC(poPreprocessing);
-     break;
-
-     case MARF.FFT:
-
      oFeatureExtraction = new
FFT(poPreprocessing);
-     break;
-
-     case MARF.F0:
-
      oFeatureExtraction = new
F0(poPreprocessing);
-     break;
-
-     case MARF.SEGMENTATION:
-
      oFeatureExtraction = new
Segmentation(poPreprocessing);
-     break;
-
-     case MARF.CEPSTRAL:
-
      oFeatureExtraction = new
Cepstral(poPreprocessing);
-     break;
-
-     case
MARF.RANDOM_FEATURE_EXTRACTION:
-
      oFeatureExtraction = new
RandomFeatureExtraction(poPreprocessing);
-     break;
-
-     case MARF.MIN_MAX_AMPLITUDES:
-
      oFeatureExtraction = new
MinMaxAmplitudes(poPreprocessing);
-     break;
-
-     case MARF.FEATURE_EXTRACTION_PLUGIN:
-     {
-     try
-           {
-
      oFeatureExtraction =
(IFeatureExtraction)MARF.getFeatureExtrac
tionPluginClass().newInstance();
-
      oFeatureExtraction.setPreprocessing
(poPreprocessing);
-                       }
-     catch(Exception e)
-                       {
-     throw new
FeatureExtractionException(e.getMessage()
, e);
-                       }
-
-     break;
-                 }
-
-     case
MARF.FEATURE_EXTRACTION_AGGREGATOR:
-                 {
-
      oFeatureExtraction = new
FeatureExtractionAggregator(poPreprocessi
ng);
-     break;
-                 }
-
-     default:
-                 {
-     throw new
FeatureExtractionException
-                       (
-  "Unknown feature extraction method: "
+ piFeatureExtractionMethod
-                       );
-                 }
-           }
+ IFeatureExtractionProvider provider =
map.get(piFeatureExtractionMethod);
+ if(provider==null)
+ throw new
FeatureExtractionException("Unknown
Feature Extraction Method: " +
piFeatureExtractionMethod);

- return oFeatureExtraction;
+ return provider.build(poPreprocessing);
      }
```

+

**Change 2 / 2: Changed implementation of create() method**
Now the key value of piFeatureExtractionMethod which is being passed as a parameter is used to fetch the relevant object from HashMap and after a null value check this object will be returned, In this way there will be no longer need of using complex switch case statement.

```
### Eclipse Workspace Patch 1.0
#P marf
Index:
src/marf/FeatureExtraction/FeatureExtract
ionFactory.java
===========================================
==========================
RCS file:
/groups/r/re_soen6471_1/cvs_repository/cv
s_repository/marf/src/marf/FeatureExtract
ion/FeatureExtractionFactory.java,v
retrieving revision 1.1.1.1
diff -u -r1.1.1.1
FeatureExtractionFactory.java
---
src/marf/FeatureExtraction/FeatureExtract
ionFactory.java   20 Aug 2014 23:55:55 -
0000  1.1.1.1
+++
src/marf/FeatureExtraction/FeatureExtract
ionFactory.java   22 Aug 2014 01:12:43 -
0000
@@ -1,5 +1,8 @@
 package marf.FeatureExtraction;

+import java.util.HashMap;
+import java.util.Map;
+
 import marf.MARF;
 import
marf.FeatureExtraction.Cepstral.Cepstral;
 import marf.FeatureExtraction.F0.F0;
@@ -65,6 +68,89 @@
  return
create(poFeatureExtractionMethod.intValue
(), poPreprocessing);
      }

+     private final static Map<Integer,
IFeatureExtractionProvider> map;
+
+     static
+     {
+       map = new HashMap<Integer,
IFeatureExtractionProvider>();
+         map.put(MARF.LPC, new
IFeatureExtractionProvider() {
+
+              @Override
+ public IFeatureExtraction
build(IPreprocessing poPreprocessing) {
+ return new LPC(poPreprocessing);
+              }
+          });
+
+ map.put(MARF.FFT, new
IFeatureExtractionProvider() {
+
+ @Override
+ public IFeatureExtraction
build(IPreprocessing poPreprocessing) {
+ return new FFT(poPreprocessing);
+              }
+          });
+ map.put(MARF.F0, new
IFeatureExtractionProvider() {
+
+ @Override
+ public IFeatureExtraction
build(IPreprocessing poPreprocessing) {
+ return new F0(poPreprocessing);
+              }
+          });
+ map.put(MARF.SEGMENTATION, new
IFeatureExtractionProvider() {
+
+ @Override
+ public IFeatureExtraction
build(IPreprocessing poPreprocessing) {
+ return new
Segmentation(poPreprocessing);
+              }
+          });
+ map.put(MARF.CEPSTRAL, new
IFeatureExtractionProvider() {
+
+ @Override
+ public IFeatureExtraction
build(IPreprocessing poPreprocessing) {
+ return new Cepstral(poPreprocessing);
+              }
+          });
+
map.put(MARF.RANDOM_FEATURE_EXTRACTION,
new IFeatureExtractionProvider() {
+
+ @Override
+ public IFeatureExtraction
build(IPreprocessing poPreprocessing) {
+ return new
RandomFeatureExtraction(poPreprocessing);
+              }
+          });
+
map.put(MARF.MIN_MAX_AMPLITUDES,       new
IFeatureExtractionProvider() {
+
```

```
+ @Override
+ public IFeatureExtraction
build(IPreprocessing poPreprocessing) {
+ return new
MinMaxAmplitudes(poPreprocessing);
+              }
+          });
+
map.put(MARF.FEATURE_EXTRACTION_PLUGIN,
new IFeatureExtractionProvider() {
+
+ @Override
+ public IFeatureExtraction
build(IPreprocessing poPreprocessing)  {
+     try
+           {
+
IFeatureExtraction oFeatureExtraction =
(IFeatureExtraction)MARF.getFeatureExtrac
tionPluginClass().newInstance();
+
oFeatureExtraction.setPreprocessing(poPre
processing);
+ return oFeatureExtraction;
+                        }
+     catch (Exception e)
+                        {
+//   throw new
FeatureExtractionException(e.getMessage()
, e);
+     return null;
+                        }
+              }
+          });
+
map.put(MARF.FEATURE_EXTRACTION_AGGREGATO
R, new IFeatureExtractionProvider() {
+
+     @Override
+     public IFeatureExtraction
build(IPreprocessing poPreprocessing) {
+     return new
FeatureExtractionAggregator(poPreprocessi
ng);
+              }
+          });
+
+     }
```

*b) Refactoring 2:*

**Change 0/2: Move method serialize() to CompilerError**

A method named serialize() was duplicated in classes LexicalError, SemanticError and SyntaxError and all of these classes have the same parent class named CompilerError, in order to remove this duplication we have moved this method to their parent class so that all of the child classes will have access to it without code redundancy.

```
-     public boolean serialize(int
piOperation, FileWriter poWriter)
-     {
-           if(piOperation == 0)
-           {
-                 // TODO load
-
      System.err.println("LexicalError::s
erialize(LOAD) - unimplemented");
-                 return false;
-           }
-           else
-           {
-                 try
-                 {
-                       poWriter.write
-                       (
-                             "Lexical
Error (line " + this.iLineNo + "): " +
this.iCurrentErrorCode +
-                             " - " +
this.strMessage + ", " +
-                             "faulting
token: [" +
this.oFaultingToken.getLexeme() + "]\n"
-                       );
-
-                       return true;
-                 }
-                 catch(IOException e)
-                 {
-
      System.err.println("LexicalError::s
erialize() - " + e.getMessage());
-                       return false;
-                 }
-           }
-     }
```

**Change 1/2: Move field oFaultingToken to CompilerError**

The Field oFaultingToken is being used by method serialize() in classes LexicalError, SemanticError and SyntaxError so after moving the method to their parent class this field is also need to be moved their.

```
diff -u -r1.1.1.1 SemanticError.java
---
src/marf/nlp/Parsing/SemanticError.java
      20 Aug 2014 23:56:00 -0000
      1.1.1.1
+++
src/marf/nlp/Parsing/SemanticError.java
      21 Aug 2014 23:11:43 -0000
@@ -69,13 +69,7 @@
            "Custom error message:"
      };
```

```
-       /**
-        * Token information at which Lexer
-        * encountered the error.
-        *
-        * @since October 2, 2001
-        */
-       protected Token oFaultingToken =
null;
```

**Change 2/2: Adding instance type checks in serialize() method**

The serialize() method has some statements to be printed on screen that shows the type of error occurred, hence in order to show the correct type of error we need to check the type of current instance. So we have added some conditions for this purpose and the proper String will be manipulated according to that showing correct type of error.

```
+       public boolean serialize(int
piOperation, FileWriter poWriter) {
+               if (piOperation == 0) {
+                       // TODO load
+                       if (this instanceof
LexicalError)
+                               System.err
+
        .println("LexicalError::serialize(L
OAD) - unimplemented");
+                       else if (this
instanceof SemanticError)
+                               System.err
+
        .println("SemanticError::serialize(
LOAD) - unimplemented");
+                       else if (this
instanceof SyntaxError)
+                               System.err
+
        .println("SyntaxError::serialize(LO
AD) - unimplemented");
+                       return false;
+               } else {
+                       try {
+                               if (this
instanceof LexicalError)
+
        poWriter.write("Lexical Error (line
" + this.iLineNo
+
        + "): " + this.iCurrentErrorCode +
" - "
+
        + this.strMessage + ", " +
"faulting token: ["
+
        + this.oFaultingToken.getLexeme() +
"]\n");
+                               else if (this
instanceof SemanticError)
+
        poWriter.write("Semantic Error
(line " + this.iLineNo
+
        + "): " + this.iCurrentErrorCode +
" - "
+
        + this.strMessage + ", " +
"faulting token: ["
+
        + this.oFaultingToken.getLexeme() +
"]\n");
+                               else if (this
instanceof SyntaxError)
+
        poWriter.write("Syntax Error (line
" + this.iLineNo + "): "
+
        + this.iCurrentErrorCode + " - " +
this.strMessage
+
        + ", " + "faulting token: ["
+
        + this.oFaultingToken.getLexeme() +
"]\n");
+
+                               return true;
+                       } catch (IOException e)
{
+
        System.err.println("CompilerError::
serialize() - "
+                                       +
e.getMessage());
+                               return false;
+                       }
+               }
+       }
```

## B. GIPSY

### 1) *Refactoring 1:*

**Change 0/1: Remove methods eval(), getBack() and execute() from class LegacyInterpreter**

The methods eval(), getBack() and execute() were duplicated in classes Interpreter and LegacyInterpreter causing code duplication in order to remove this code smell we have removed these methods from LegacyInterpreter.

```
-       public GIPSYContext[]
getBack(GIPSYContext[]
paoOriginalContext)
-       {
-               GIPSYContext[] aoContextTmp =
new GIPSYContext[DIMENSION_MAX];
-               Arrays.copy(aoContextTmp,
paoOriginalContext, DIMENSION_MAX);
-               return aoContextTmp;
-       }
```

```
-
-     /**
       * Older implementation of the main
evaluation routine
       * based on the local interpreter,
integer dimensions,
       * and integer types.
@@ -525,318 +514,6 @@

          throw new
GIPSYRuntimeException();
      }
-
-     /**
-      * New version of
<code>eval()</code> that corresponds to
the new definitions
-      * of GIPSYContext.
-      *
-      * @param poRoot root of the AST
-      * @param paoContext
-      * @param piIndent
-      * @return
-      * @throws GEEException
-      *
-      * @since Serguei Mokhov
-      */
-     public GIPSYType eval(SimpleNode
poRoot, GIPSYContext[] paoContext, int
piIndent)
-     throws GEEException
-     {
-           //int iResult1, iResult2;
-           int iDimension;
-
-
      //IArithmeticOperatorsProvider
oResult1;
-           // XXX: pull up to a instance
or class instance
-           IArithmeticOperatorsProvider
oArithmeticAlgebra = new
GenericArithmeticOperatorsDelegate();
-
-           GIPSYType oResult1;
-           GIPSYType oResult2;
-
-           // just to keep the original
context
-           GIPSYContext[]
aoOriginalContext = new
GIPSYContext[DIMENSION_MAX];
-
      Arrays.copy(aoOriginalContext, 0,
paoContext);
-
-
-           // The order of children is
fixed in the syntax checking
-           switch(poRoot.id)
-           {
-                 // @ d
-                 case JJTAT:
-                 {
-                       iDimension = (
(SimpleNode) poRoot.children[1] ).ID ;
-
      paoContext[iDimension] =
(GIPSYContext)eval( (SimpleNode)
poRoot.children[2], paoContext, piIndent
+ 1);
-                       return eval(
(SimpleNode) poRoot.children[0],
paoContext, piIndent + 1);
-                 }
-
-                 // # d
-                 case JJTHASH:
-                 {
-                       iDimension = (
(SimpleNode) poRoot.children[0] ).ID ;
-                       return
paoContext[iDimension];
-                 }
-
-                 // arith_op
-
-                 // x + y
-                 case JJTADD:
-                 {
-                       oResult1 =
(GIPSYType)eval((SimpleNode)poRoot.childr
en[0], paoContext, piIndent + 1);
-                       paoContext =
getBack(aoOriginalContext);
-                       oResult2 =
eval((SimpleNode) poRoot.children[1],
paoContext, piIndent + 1);
-                       return
oArithmeticAlgebra.add(oResult1,
oResult2);
-                 }
-
-                 // x - y
-                 case JJTMIN:
-                 {
-                       oResult1 =
(GIPSYInteger)eval((SimpleNode)poRoot.chi
ldren[0], paoContext, piIndent + 1);
-                       paoContext =
getBack(aoOriginalContext);
-                       oResult2 =
eval((SimpleNode)poRoot.children[1],
paoContext, piIndent + 1);
```

```
-                        //return
((GIPSYInteger)oResult1).subtract(oResult
2);
-                        return
oArithmeticAlgebra.subtract(oResult1,
oResult2);
-                  }
-/*
-         case JJTTIMES:
-           iResult1 = eval((SimpleNode)
oRoot.children[0], paoContext, piIndent +
1) ;
-           paoContext = getBack(
aoOriginalContext );
-           iResult2 = eval((SimpleNode)
oRoot.children[1], paoContext, piIndent +
1) ;
-           return iResult1 * iResult2;
-
-         case JJTDIV:
-           iResult1 = eval((SimpleNode)
oRoot.children[0], paoContext, piIndent +
1) ;
-           paoContext = getBack(
aoOriginalContext );
-           iResult2 = eval((SimpleNode)
oRoot.children[1], paoContext, piIndent +
1) ;
-           return iResult1 / iResult2;
-
-         case JJTMOD:
-           iResult1 = eval((SimpleNode)
oRoot.children[0], paoContext, piIndent +
1) ;
-           paoContext = getBack(
aoOriginalContext );
-           iResult2 = eval((SimpleNode)
oRoot.children[1], paoContext, piIndent +
1) ;
-           return iResult1 % iResult2;
-
-           // rel_op
-         case JJTLT:
-           iResult1 = eval((SimpleNode)
oRoot.children[0], paoContext, piIndent +
1) ;
-           paoContext = getBack(
aoOriginalContext );
-           iResult2 = eval((SimpleNode)
oRoot.children[1], paoContext, piIndent +
1) ;
-
-            if(iResult1 < iResult2)
-                        return 1;
-                  else
-                        return 0;
-
-         case JJTGT:
-           iResult1 = eval((SimpleNode)
oRoot.children[0], paoContext, piIndent +
1) ;
-           paoContext = getBack(
aoOriginalContext );
-           iResult2 = eval((SimpleNode)
oRoot.children[1], paoContext, piIndent +
1) ;
-           if (iResult1 > iResult2)
return 1;
-           else return 0;
-
-         case JJTLE:
-           iResult1 = eval((SimpleNode)
oRoot.children[0], paoContext, piIndent +
1) ;
-           paoContext = getBack(
aoOriginalContext );
-           iResult2 = eval((SimpleNode)
oRoot.children[1], paoContext, piIndent +
1) ;
-           if (iResult1 <= iResult2)
return 1;
-           else return 0;
-
-       case JJTGE:
-           iResult1 = eval((SimpleNode)
oRoot.children[0], paoContext, piIndent +
1) ;
-           paoContext = getBack(
aoOriginalContext );
-           iResult2 = eval((SimpleNode)
oRoot.children[1], paoContext, piIndent +
1) ;
-           if (iResult1 >= iResult2)
return 1;
-           else return 0;
-
-         case JJTEQ:
-           iResult1 = eval((SimpleNode)
oRoot.children[0], paoContext, piIndent +
1) ;
-           paoContext = getBack(
aoOriginalContext );
-           iResult2 = eval((SimpleNode)
oRoot.children[1], paoContext, piIndent +
1) ;
-           if (iResult1 == iResult2)
return 1;
-           else return 0;
-
-         case JJTNE:
-           iResult1 = eval((SimpleNode)
oRoot.children[0], paoContext, piIndent +
1) ;
-           paoContext = getBack(
aoOriginalContext );
```

```
-          iResult2 = eval((SimpleNode)
oRoot.children[1], paoContext, piIndent +
1) ;
-          if (iResult1 != iResult2)
return 1;
-          else return 0;
-
-// log_op
-        case JJTAND:
-          iResult1 = eval((SimpleNode)
oRoot.children[0], paoContext, piIndent +
1) ;
-          paoContext = getBack(
aoOriginalContext );
-          iResult2 = eval((SimpleNode)
oRoot.children[1], paoContext, piIndent +
1) ;
-          if ( iResult1 == 1 && iResult2
== 1 ) return 1;
-          else return 0;
-
-        case JJTOR:
-          iResult1 = eval((SimpleNode)
oRoot.children[0], paoContext, piIndent +
1) ;
-          paoContext = getBack(
aoOriginalContext );
-          iResult2 = eval((SimpleNode)
oRoot.children[1], paoContext, piIndent +
1) ;
-          if ( iResult1 == 1 || iResult2
== 1 ) return 1;  // true
-          else return 0;    // false
-
-        case JJTIF:
-          if (eval((SimpleNode)
oRoot.children[0], paoContext, piIndent +
1) == 1 )   //if true
-          {
-             paoContext = getBack(
aoOriginalContext );
-             return eval((SimpleNode)
oRoot.children[1], paoContext, piIndent +
1);
-          }
-          else
-          {
-             paoContext = getBack(
aoOriginalContext );
-             return eval((SimpleNode)
oRoot.children[2], paoContext, piIndent +
1);
-          }
-
-// The key of the vhouse is a string
combined by the id and its context
-// Enquire the id+context in the value
house
-//   if it exists, return the value
-//   if it doesn't, evaluate the
id+context
-
-        case JJTID:
-          String s = ( new Integer
(oRoot.ID) ).toString();
-
-          try
-          {
-            DictionaryItem identry =
(DictionaryItem) oDictionary.get(
oRoot.ID );
-            StringBuffer sbuff = new
StringBuffer(s);
-
-            for ( int i = 0 ; i <
DIMENSION_MAX; i++)
-            {
-              sbuff = sbuff.append(",");
-              Integer tmp = new Integer
(paoContext[i]);
-              sbuff = sbuff.append
(tmp.toString());
-            }
-
-            //Integer result =
(Integer)vhouse.get(sbuff.toString());
-            Integer result =
(Integer)oValueHouse.getValue(sbuff.toStr
ing());
-
-            if(result == null)
-            {
-              int result_eval =
eval((SimpleNode)(identry.getEntry()),
paoContext, piIndent + 1);
-
//vhouse.put(sbuff.toString(),new
Integer(result_eval));
-
oValueHouse.setValue(sbuff.toString(),new
Integer(result_eval));
-              return result_eval;
-            }
-            else
-            {
-              return result.intValue();
-            }
-
-          }
-
catch(ArrayIndexOutOfBoundsException e)
// in case the id is null
-          {
-
//System.err.println("LegacyInterpreter
```

```
couldn't resolve the identifier symbol: "
+ s);
-            throw new
GEEException("LegacyInterpreter couldn't
resolve the identifier symbol: " + s);
-          }
-
-          // the value of a constant is
its image
-        case JJTCONST:
-        case
IndexicalLucidParserTreeConstants.JJTCONS
T:
-        {
-                 return ( new Integer (
oRoot.getImage() ) ).intValue();
-          //return ( new Number (
expr.getImage() ) ).intValue();
-        }
-
-          */
-        default:
-         throw new
GEEException("LegacyInterpreter bad node
type: " + poRoot.id);
-    }
-  }
-
-     /**
-      * Main execution method.
-      * @param poDictionary
-      * @param poDimensionTags
-      */
-     public GIPSYType execute(Dictionary
poDictionary, GIPSYContext
poDimensionTags)
-     {
-           DictionaryItem oItem;
-
-           boolean bFini = false;
-
-           for(int k = 0; k <
poDictionary.size(); k++)
-           {
-                 oItem =
(DictionaryItem)poDictionary.elementAt(k)
;
-
      this.oDictionary.add(oItem.getID(),
oItem);
-
      this.oMiniDictionary.add(oItem.getI
D(), oItem.getName());
-           }
-
-           Debug.debug("pram dic: " +
poDictionary + ", this dic: " +
this.oDictionary + ", mini dic: " +
this.oMiniDictionary);
-
-           int[] aiContexts = new
int[DIMENSION_MAX];
-
-           for(int iContextualIndex = 0
; iContextualIndex < DIMENSION_MAX;
iContextualIndex++)
-           {
-
      aiContexts[iContextualIndex] = 0;
-           }
-
-           // buildDict();
-
-           // parse dimensions values,
which are in a form: d=2,m=3
-
-           //
pstrDimensionValues = "d=2";
-
-           //            StringBuffer
oDimensionValue = new
StringBuffer(poDimensionTags);
-           StringBuffer oDimensionValue
= new
StringBuffer(poDimensionTags.toString());
-           oDimensionValue.append(",");
-
-           String subS, subS1, subS2;
-
-           int idx, idx2;
-           int dimId;
-
-
      while(oDimensionValue.length() !=
0)
-           {
-                 // TODO: this parsing
business does not belong
-                 // to the interpreter.
Ought to move it to GIPC.
-                 idx =
oDimensionValue.indexOf(",");
-                 subS =
oDimensionValue.substring(0, idx);
-
-                 subS1 = new
String(subS);
-                 subS2 = new
String(subS);
-                 idx2 =
subS.indexOf("=");
-                 subS2 =
subS1.substring(idx2 + 1);
-                 subS1 =
subS.substring(0, idx2);
```

```
-                  dimId =
oMiniDictionary.indexOf(subS1);
-
-
      System.out.println("idx="+idx+",
subS="+subS+", subS1="+subS1+",
subS2="+subS2+",idx2="+idx2+",dimId="+dim
Id);
-
      System.out.println("cont.length="+a
iContexts.length);
-
-                  //TODO: need to guard
dimId against DIMENSION_MAX (eg.
DIMENSION_MAX = 1 causes this to fail
-                  if(dimId != -1)
-                  {
-                        aiContexts[dimId]
= (new Integer(subS2)).intValue();
-                  }
-                  else
-                  {
-
      System.err.println("LegacyInterpret
er cannot resolve dimension symbol: " +
subS1);
-                        bFini = true;
// if the demanded thing can't be
resolved, no further process
-                  }
-
-                  oDimensionValue = new
StringBuffer(oDimensionValue.substring(id
x + 1));
-            }
-
-            if(!bFini)
-            {
-                  try
-                  {
-                        oItem =
(DictionaryItem)oDictionary.get(2);
-
      Debug.debug("LegacyInterpreter
item: " + oItem);
-                        //int result =
eval((SimpleNode)(oItem.getEntry()),
aiContexts, 0);
-                        GIPSYType oReuslt
= eval((SimpleNode)(oItem.getEntry()),
new GIPSYContext[] {poDimensionTags}, 0);
-
      System.out.println("The result of
calculation is: " + oReuslt);
-                        return oReuslt;
-                  }
-                  catch(Exception e)
-                  {
-
      System.err.println("LegacyInterpret
er error calculating the result.");
-
      System.err.println(e.getMessage());
-
      e.printStackTrace(System.err);
-                        throw new
GIPSYRuntimeException(e);
-                  }
-            }
-
-            throw new
GIPSYRuntimeException();
-      }
```

**Change 1/1: Change parent class of LegacyInterpreter**
As the methods eval(), getBack() and execute() are already available in Interpreter class and they have been removed from here, hence we have changed the parent class of LegacyInterpreter to Interpreter so that these methods are still available to it without code redundancy.

```
Index:
src/gipsy/GEE/IDP/DemandGenerator/LegacyI
nterpreter.java
=========================================
==========================
RCS file:
/groups/r/re_soen6471_1/cvs_repository/cv
s_repository/gipsy/src/gipsy/GEE/IDP/Dema
ndGenerator/LegacyInterpreter.java,v
retrieving revision 1.1.1.1
diff -u -r1.1.1.1 LegacyInterpreter.java
---
src/gipsy/GEE/IDP/DemandGenerator/LegacyI
nterpreter.java   20 Aug 2014 23:56:11 -
0000  1.1.1.1
+++
src/gipsy/GEE/IDP/DemandGenerator/LegacyI
nterpreter.java   22 Aug 2014 00:39:41 -
0000
@@ -35,7 +35,7 @@
  * @since November 18, 2002
  */
 public class LegacyInterpreter
-extends DemandGenerator
+extends Interpreter
 implements CONFIG,
GIPLParserTreeConstants
```

## VIII. CONCLUSION:

The current software world is dynamic and inconsistent. Its facing challenges and changes day to day, hour to hour, minute to minute and second to second. Refactoring is essential in any software project. We remember “No Silver Bullet”. Refactoring is the tool used for several purposes. Poorly designed code need refactoring. Legacy code also requires refactoring. Refactoring is used to improve the code but not performance and makes software easier to understand. Refactoring helps to find the bugs in a program and a part of Reverse Engineering. Refactoring is done based on three temporal based factors. When a new function is added, during code review, and while fixing a bug. Patterns are widely recognized in the interaction design and Human Computer Interaction and also software development. GIPSY is a framework for compilation of heterogeneous intensional programs. DMARF is distributed used for Forensic analysis. Initially, Logiscope was used for the metrics calculation of classes, Java Files in both the DMARF and GIPSY frameworks. We learned the concepts of Refactoring and Patterns by doing Refactoring on GIPSY and DMARF. Identifying code smells is the pre-procedure for refactoring and we have identified the code smells using JDeodrant. The refactoring was done on the identified classes. Duplicate classes were identified and defined an interface for them in refactoring. We used objectAid for the reverse engineering and for the patterns. It is an eclipse plugin and used for building the classes, sequence diagrams. Microsoft Visio made an easy way to build the class diagrams from the readings of the GIPSY and DMARF case studies. CVS is a tool used for developing projects in a group environment. It provides version control system and at any time the previous commits, changes can be known and recovered. Its very helpful in developing projects that can be revoverable, changeable. Finally, the concepts of Metrics, Refactoring, Patters, CVS repository made us to learn the real time project development and reverse engineering.

## IX. Glossary

1. **CORBA** Common Object Request Broker Architecture – a language model independent platform for distributed execution of applications possibly written in different languages, and, is, therefore, heterogeneous type of RPC (unlike Java RMI, which is Java-specific).

2. **RMI** Remote Method Invocation - an object-oriented way of calling methods of objects possibly stored remotely with respect to the calling program.

3. **GMT** GIPSY Manager Tier - GIPSY managers are instances of this tier which enable registration of a GIPSY node to a GIPSY instance and allocation of various GIPSY tiers to these nodes.

4. **DST** Demand Store Tier - It is a middle ware tier that exposes other tiers to GIPSY Demand Migration System (DMS). This is a communication system that connects tiers through demands.

5. **DGT** Demand Generator Tier - Instances of this tier encapsulate in itself a demand generator. The purpose of this is to generate demands.

6. **DWT** Demand Worker Tier - Instances of this tier contain a set of demand workers that process procedural demands resulting in a compiled version.